UNIVERSIDADE DE LISBOA
FACULDADE DE CIÊNCIAS
DEPARTAMENTO DE FÍSICA

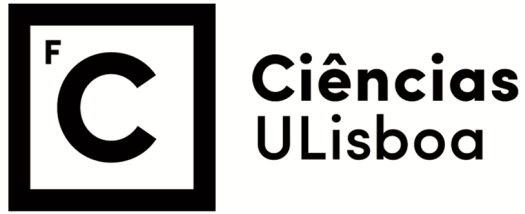

# Constraining f(Q) Cosmology with Standard Sirens

José Pedro Mota Valente Ferreira

**Mestrado em Física**
Especialização em Astrofísica e Cosmologia

Dissertação orientada por:
Prof. Doutor Nelson Nunes
Prof. Doutor Tiago Barreiro

2022

*Já são dez para as seis da manhã*
*E há mais um imbecil a pensar*
*Este mundo não foi feito para a gente entender*
*Nem por isso eu deixo de o tentar*

---

Manuel Cruz
*Tirem o macaco da prisão*

iv

# Agradecimentos

Mesmo sendo, para todos os efeitos, os co-autores desta tese, devo o meu agradecimento ao Nelson e ao Tiago, por me terem orientado ao longo deste que foi o meu primeiro ano no mundo da investigação. Sei, certamente, muito mais hoje do que quando comecei esta demanda. Naturalmente, este agradecimento também se estende ao José Mimoso, ocasionalmente mencionado como *Zé Pedro I*, pelo papel desempenhado como para-orientador. As suas principais contribuições incluem a quantidade incomensurável de ideias e divagações, que passam pelos vários cantos do mundo da física, bem como o das suas principais personagens (nacionais e internacionais, presentes e passadas). Foi certamente graças aos três que este primeiro ano no mundo da investigação foi tão produtivo e interessante.

Tenho também que agradecer à minha família, no qual incluo, naturalmente, a Clara, pelo apoio prestado, bem como as permanentes tentativas de tentar decifrar as coisas que vou dizendo, ocasionalmente incompreensíveis para quem as ouve, incluindo para o próprio.

Por fim, gostava de expressar o meu agradecimento ao Diogo Santos, o meu camarada, de jogo, que me conseguiu ensinar mais sobre estatística e economia do que eu lhe consegui ensinar física e GNU/Linux.

Assinado,
*Zé Pedro II*

# **Resumo**

A nossa dissertação tem como objetivo estudar dois modelos cosmológicos, que têm como base teorias de gravidade modificadas que envolvem à não-metricidade, em vez da tradicional teoria da Relatividade Geral. Recorrendo a catálogos simulados de sirenes padrão, vamos prever a capacidade de futuros observatórios de ondas gravitacionais conseguirem constranger os parâmetros presentes nesses mesmos modelos, bem como se serão capazes de os distinguir do modelo padrão da cosmologia.

O modelo padrão da cosmologia, conhecido como $\Lambda$CDM, foi desenvolvido ao longo do último século com o intuito de descrever o Universo na sua maior escala. Sendo relativamente simples na sua formulação teórica, tendo por base um universo isotrópico e homogéneo com interações descritas pela teoria da Relatividade Geral, tem sido validado nas últimas duas décadas através das suas previsões, face a observações feitas com cada vez maior precisão. No entanto, apesar dos seus grandes sucessos, existem ainda várias questões por esclarecer. Em particular, a existência de observações em tensão entre si, das quais se destaca a tensão de Hubble, a anomalia na radiação cósmica de fundo, bem como a falta de deteção em laboratório dos dois componentes mais prevalecentes do universo, a matéria e energia escura.

Existem, naturalmente, inúmeras abordagens possíveis para tentar encontrar um novo modelo que seja capaz de resolver os problemas que existem em torno de $\Lambda$CDM. Nesta dissertação, vamos considerar teorias de gravidade modificada baseadas em não-metricidade, que foram inicialmente propostas com o intuito de unificar a gravitação com o eletromagnetismo, e vão ser agora utilizadas para construir modelos cosmológicos alternativos ao modelo padrão da cosmologia.

A Relatividade Geral como a descrição da interação gravitacional teve como consequência a previsão de vários fenómenos até então inobserváveis. Um desses fenómenos, que tem particular interesse para esta dissertação, é a existência de ondas gravitacionais, que foram observadas pela primeira vez por uma análise indireta do sistema binário de Hulse-Taylor em 1974, e mais tarde, em 2015, medidas diretamente pelo Laser Interferometer Gravitational-Wave Observatory (LIGO). Estes eventos têm como principal origem sistemas binários de corpos extremamente compactos (por exemplo buracos negros, estrelas de neutrões). Estes eventos são possíveis pois na teoria da Relatividade Geral os sistemas binários são instáveis, emitindo ondas gravitacionais gradualmente, fazendo com que a distância entre os corpos diminua, até à sua colisão. No caso em que os corpos compactos emitem também ondas eletromagnéticas, como por exemplo no caso da colisão de duas estrelas de neutrões, temos um evento que se chama sirene padrão. Estes eventos são de extrema importância para o estudo da cosmologia, pois é possível obter o desvio para o vermelho a partir da onda eletromagnética e a distância luminosa a partir da onda gravitacional. Isto permite-nos constranger um modelo cosmológico sem recorrer a uma escada de distâncias cosmológicas, evitando assim eventuais erros de calibração.

Até à data apenas houve uma ocorrência confirmada de um evento deste tipo, o GW170817. Não é portanto possível constranger modelos com dados actuais de sirenes padrão, e vamos alternativamente, criar catálogos simulados de sirenes padrão, com o intuito de conseguir saber se no futuro será possível distinguir os modelos cosmológicos que vamos considerar ao longo desta dissertação do modelo padrão da cosmologia.

No primeiro capítulo desta dissertação vamos fazer uma introdução geral aos conceitos que vamos desenvolver ao longo do documento e que vão também servir de motivação para o trabalho desenvolvido. Iremos destacar qual o objetivo principal da cosmologia como área de estudo, algumas das características e falhas de $\Lambda$CDM, o facto de ser possível ver a gravidade como uma consequência da geometria do espaço-tempo e uma breve introdução a sirenes padrão. Os principais objetivos do trabalho, bem como a estrutura do documento, também estarão presentes nesse mesmo capítulo.

No segundo capítulo, iremos apresentar a gravidade como a consequência da geometria do espaço-tempo. Vamos também mostrar que, ao contrário do que acontece na teoria da Relatividade Geral, não é necessária a existência de curvatura para descrever esta interação. A fim de poder fazer essa construção, vamos introduzir brevemente o conceito de não-metricidade e de torção, dois objetos que são assumidos ser nulos na teoria da Relatividade Geral. Com estes objetos vamos mostrar que é possível construir três teorias equivalentes entre si, na qual uma delas é uma teoria exclusivamente não métrica, baseada no escalar da não-metricidade $Q$, a outra é baseada apenas em torção, e a terceira é a teoria da Relatividade Geral, que depende unicamente da curvatura. De seguida vamos generalizar cada uma destas teorias promovendo a quantidade escalar na ação para uma função arbitrária do mesmo, com particular interesse na teoria de $f(Q)$, que estará no centro desta dissertação.

No terceiro capítulo, iremos introduzir formalmente as considerações por detrás dos nossos modelos cosmológicos. Ao longo desta dissertação foi considerado que o Universo nas grandes escalas é homogéneo e isotrópico, bem como espacialmente plano, e encontra-se preenchido por um fluido perfeito. Em seguida, vamos apresentar o modelo padrão da cosmologia e as modificações que têm lugar ao considerar modelos cosmológicos baseados em $f(Q)$, tanto na evolução do fundo cósmico como na propagação de ondas gravitacionais.

No quarto capítulo vamos expor as fontes de dados que utilizámos ao longo do trabalho, desde as Supernova tipo Ia ao processo de gerar catálogos simulados de sirenes padrão para todos os observatórios considerados. Para isso, vamos descrever as características dos detetores de ondas gravitacionais da colaboração LIGO-Virgo, bem como as expectativas existentes para futuros observatórios como o Laser Interferometer Space Antenna (LISA) e o Einstein Telescope (ET). O final deste capítulo será dedicado à metodologia da nossa análise estatística, descrevendo como e com que ferramentas foram realizados os constrangimentos dos modelos, bem como o processo de seleção de modelos e de catálogos.

O quinto capítulo será dedicado ao modelo de $f(Q)$ mais geral que replica a dinâmica de fundo de $\Lambda$CDM, apresentado diferenças somente na propagação de perturbações. Ao introduzir apenas um parâmetro adicional quando comparado com ΛCDM, denominado por $\alpha$, iremos analisar qual é a capacidade de cada observatório de colocar limites no valor do parâmetro $\alpha$. Foi observado que o LIGO-Virgo não é capaz de colocar limites ao valor de $\alpha$, enquanto que o ET e o LISA são, com o ET a colocar limites mais fortes do que o LISA. Foi observado que tanto o LIGO-Virgo como o LISA sofrem de flutuações estatísticas nos seus catálogos, pelo que serão considerados três casos representativos: o melhor, o mediano e o pior. Pelo contrário, o ET não sofre destas flutuações estatísticas, dado que o número de eventos é suficientemente grande para que essas flutuações sejam desprezáveis. Mostrámos também que mesmo se no futuro se obtiver um mau catálogo do LISA, podemos utilizar os dados do LIGO-Virgo para aproximar esse catálogo a um catálogo mediano do LISA.

No sexto capítulo iremos estudar um modelo de $f(Q)$ que tem como objetivo substituir a existência de energia escura através da introdução de uma forma específica na modificação à gravidade. Em primeira instância foi feita uma análise recorrendo a técnicas de sistemas dinâmicos para analisar regiões no espaço de parâmetros que apresentam cosmologias que estão de acordo com as observações atuais. Utilizando o processo de seleção de modelos introduzido previamente, recorrendo apenas a sirenes padrão, é impossível distinguir este modelo de ΛCDM. No entanto, ao introduzirmos os dados de Supernovas do tipo Ia, surgem tensões no modelo quando comparado com sirenes padrão. Significa isto que, no futuro, será possível utilizar dados de sirenes padrão para distinguir este modelo de ΛCDM.

No sétimo e último capítulo encerramos a dissertação fazendo considerações finais, bem como apresentar perspetivas futuras para dar continuidade ao trabalho aqui desenvolvido.

**Palavras-chave:** Sirenes Padrão, Ondas Gravitacionais, Astronomia Multimensageira, Gravidade Modificada, Cosmologia Observacional

# Abstract

In this dissertation, we study two cosmological models based on $f(Q)$ gravity. We resort to mock catalogs of standard siren (SS) events to see whether data from future GW observatories will be able to distinguish these models from $\Lambda$CDM.

The first model is the most general $f(Q)$ formulation that replicates a $\Lambda$CDM background, with deviations appearing only at the perturbative level. It has one additional free parameter compared to $\Lambda$CDM, $\alpha$, which when set to zero falls back to $\Lambda$CDM. We show that LIGO-Virgo is unable to constrain $\alpha$, due to the high error and low redshift of the measurements, whereas LISA and the ET will, with the ET outperforming LISA. The catalogs for both LISA and LIGO-Virgo show non-negligible statistical fluctuations, where we consider three representative catalogs (the best, median and worst), whereas for the ET, only a single catalog is considered, as the number of events is large enough for statistical fluctuations to be neglected. The best LISA catalog is the one with more low redshift events, while the worst LISA catalog features fewer low redshift events. Additionally, if we are to observe a bad LISA catalog, we can rely on data from LIGO-Virgo to improve the quality of the constrains, bringing it closer to a median LISA catalog.

The second model attempts to replace dark energy by making use of a specific form of the function $f(Q)$. We study this model resorting to dynamical system techniques to show the regions in parameter space with viable cosmologies. Using model selection criteria, we show that no number of SS events is, by itself, able to tell this model and $\Lambda$CDM apart. We then show that if we add current SnIa data, tensions in this model arise when compared to the constrains set by the SS events.

**Keywords:** Multi-messeger Astronomy, Modified Gravity, Observational Cosmology, Standard Sirens, Gravitational Waves

# Contents

# List of Tables

# List of Figures

# Acronyms

**ΛCDM**  Λ Cold Dark Matter  vii–ix, xi, 2–5, 15, 17, 19, 20, 23, 26, 33–39, 41, 44, 46–54

**BNS**  binary neutron star  25

**CDM**  cold dark matter  44, 45, 52, 53

**CMB**  cosmic microwave background  2, 3, 54

**EFE**  Einstein Field Equations  11, 12, 17, 19

**elpd**  expected log pointwise predictive density  30, 31, 47–51

**EM**  electromagnetic  4, 20, 21, 25

**EoS**  equation of state  16, 17

**ET**  Einstein Telescope  viii, ix, xi, 25, 29, 35–40, 47–49, 51–53

**FLRW**  Friedmann Lemaître Robertson Walker  15, 17, 18, 20

**GR**  General Relativity  2–4, 7, 9–12, 17–20, 25, 52

**GW**  gravitational wave  xi, 4, 19–21, 23, 25, 28, 33, 34, 47, 53, 54

**LIGO**  Laser Interferometer Gravitational-Wave Observatory  vii–ix, xi, 25, 26, 36–40, 48, 53

**LISA**  Laser Interferometer Space Antenna  viii, ix, xi, xvii, 25, 27, 28, 34–40, 47–49, 51, 53

**LOO-CV**  leave-one-out cross-validation  31

**MBHB**  massive black hole binary  25, 27–29

**MCMC**  Markov chain Monte Carlo  30, 31

**PSIS-LOO-CV**  Pareto smoothed importance sampling leave-one-out cross-validation  31, 47–51

**SnIa**  type Ia supernova  xi, 2, 5, 23, 24, 26, 34–36, 38, 39, 41, 50–54

**SS**  standard siren  xi, 4, 5, 19, 23, 25–27, 29, 32–34, 39, 41, 47, 48, 50–54

**STEGR** Symmetric Teleparallel Equivalent of General Relativity 12, 13, 44, 52

**STG** Symmetric Teleparallel Gravity 13

**TEGR** Teleparallel Equivalent of General Relativity 12, 13

**TG** Teleparallel Gravity 13

# Chapter 1

# Introduction

In this chapter, we will give a broad overview of the topics which are at the heart of this dissertation, aiming to give the reader a brief introduction on the subjects to come, which will also serve as motivation for the work carried out. In the two final sections of this chapter, we will outline the main goals for this dissertation and the structure of the document.

## 1.1 Cosmology: The Study of the Cosmos

Cosmology is the branch of physics concerned with the study of the Universe at the largest scales. Cosmology knows no physical bounds, neither in space, nor in time, taking the challenge of modeling the evolution of the Universe throughout time.

To encompass the evolution of the Universe as a whole is no easy feat, given that throughout history there were no lack of attempts. Perhaps one of the most interesting of such attempts is the Aristotelian-Ptolemaic cosmological model, one of the earliest cosmological theories that we have record of. With roots in ancient Greece, this paradigm stated that the Universe is made of spherical shells consisting of the planets and stars, as well as a handful of very primitive elements (earth, water, fire and air). Although we regard this today as an insufficient approximation of reality, worthy of a mythological tale rather than a scientific theory, it is considered until this day to be the most long-lived scientific paradigm in history [1].

We might ask ourselves, how was such a naive model able to withstand for so long as a good description of the cosmos? After all, the Universe that we know exists out there differs wildly from these ideas. The reason as to why this model lived for so long was simply because the tools required to see what the Universe really look like were only developed in the centuries to come. Today, with an unprecedented amount of high quality data available to us, we have to be ever more sophisticated in order to accurately predict observations across a wide range of phenomena. Given the importance that new measurements have on the development of our ideas, and the finite number of resources we have available, choosing our investments carefully is a requirement in order to expand our knowledge as fast at we can.

## 1.2 The Standard Cosmological Model

In our attempts to contain the evolution of the cosmos in a single theory, the standard model of cosmology, referred to as ΛCDM, was developed. Starting its roots in the 20th century, it is till today the dominant paradigm in modern cosmology and while remaining fairly simple in its theoretical formulation, is able to successfully account for most of the observed phenomena.

One of the requirements in order to understand the evolution of the cosmos, is the ability to describe the motion of bodies when acted upon by gravity. Although there are several theories of gravity which one can pick from, Einstein's theory of General Relativity (GR) is undoubtedly the most successful one so far. Since the formulation of GR, we see the motion of a particle in a gravitational field as the consequence of the curvature of spacetime, a four dimensional manifold where space and time are fused together into one mathematical object. Thus, knowing how spacetime curves, which can be computed knowing just how much matter content there is, means that one can make predictions regarding the future state of any given particle. Initially starting with the prediction of the perihelion precession of Mercury and the deflection of light rays, when the Universe was thought to be our galaxy and a background of stars, it successfully accounts for a wide range of phenomena which extended far beyond the known Universe at that time. For all of its predictions and agreement with experiment, this theory is, therefore, at the heart of the standard cosmological model.

While during the early years of GR there were few observations of the Universe at cosmological scales, this rapidly changed due to the major technological developments that followed. Perhaps a few of the more surprising observations were the detection of the excessive speed of the galaxies and missing mass observed in the Coma galaxy cluster, by Zwicky in 1933 [2], and in 1998 where measurements from type Ia supernova (SnIa) were used to show an accelerating expanding Universe [3]. These observations lead to the concept of dark matter and dark energy respectively, which are known today to be a major portion of our Universe. The other part of our Universe consists of less exotic components, which surround us in our everyday life, which we differentiate between ordinary matter and radiation.

Given that in GR one requires the knowledge of the matter distribution in order to compute the geometry of the Universe, we are lead, by observations, to the idea that at sufficiently large scales, the Universe is homogeneous (i.e. no point is privileged when compared to the other) and isotropic (i.e. at a given point, the Universe looks the same in all directions). This is known as the cosmological principle, and is one of the main ideas behind the standard cosmological model. Recent observations have lead the community to favor a universe which is spatially flat.

With this surprisingly straightforward theoretical formulation, this model as been able to accurately account for many of the observed phenomena at largest scale, of which the most prominent are [4]:

1. The existence of the Cosmic Microwave Background (CMB);

2. The accelerated expansion of the Universe;

3. The abundances of the light elements;

4. The large scale distribution of galaxies.

However, with respect to those same observations, ΛCDM is currently facing some observational challenges. As listed in [5], the most significant are:

1. **The Hubble crisis**: A tension between the measurements of the Hubble constant between low and high redshift observations;

2. **Anomalies in the anisotropies of CMB**: Unexplained distribution of the temperature fluctuations in CMB which are not accounted for in ΛCDM;

3. **The $\sigma_8$ tension**: A tension on the amplitude of the matter power spectrum at the scale $8h^{-1}$ Mpc between large scale structure data and Plack measurements.

Additionally, one could also list other problems, such as the lack of theoretical explanation of the two most prevalent components of our Universe, dark energy and dark matter. The standard model of particle physics fails to predict the energy density value of the dark energy by a factor of $10^{120}$ [6]. Also the incompatibility of GR with well established microscopic theories, which was the driving source of evolution for several modifications to gravity in starting the past century.

## 1.3  Beyond the Standard Model

Faced with the issues we have just discussed in the previous section, one should seek new models that might improve on ΛCDM. There is a countless number of ways to modify ΛCDM, in this work, we have decided to explore possible modifications to GR, the theory of gravity at the core of ΛCDM.

Einstein initially formulated GR as a geometrical theory of gravity where the gravitational phenomena are described by the curvature of spacetime. Despite being the more common approach, the geometry of spacetime might not necessarily be due to curvature but other geometrical objects, such as torsion and non-metricity. These alternative descriptions of gravity, that preserves the equivalence principle, can be used to create theories that may or may not resemble GR. The theories equivalent to GR, often referred to as different interpretations of GR, have been recently popularized in what is now known as the geometrical trinity of gravity [7].

The idea of using other geometrical objects in an attempt to explain for unaccounted phenomena by creating modified theories of gravity is by no means contemporary. One of the earliest attempts at modifying gravity was developed by Weyl, in 1919, where he used non-metricity to attempt to unify gravity and electromagnetism [8]. Later, in 1923, Cartan developed a modified theory of gravity based on torsion instead [9].

In this work we will pursue the path of modifying gravity and will rely on a non-metricity scalar, $Q$, by considering an action that depends on an arbitrary function of this quantity, $f(Q)$. We will then see whether specific forms of $f(Q)$ are able to compete against the standard cosmological model.

## 1.4 Standard Sirens

Much like we have hinted previously during this introduction, data is essential to bring any physical theory onto solid ground. In order to further test our models for both gravity and cosmology, we are now beginning an era which brings a new way to study the Universe, gravitational wave (GW) astronomy. This new way to "listen" to the Universe, has already provided us with remarkable results, and further are to be expected in the future from future ground and space based observatories which are currently being developed. With this new source of data, we will be able to further address the difficulties that ΛCDM faces, and also to study gravity in the high energy regime, allowing us to investigate deviations from GR.

A specific type of phenomena of particular interest to cosmology are standard siren (SS) events. They are characterized by the emission of both GWs and electromagnetic (EM) radiation, which are expected to take place in the merger of binary system of compact objects, due to the decrease of the orbital radius by slowly loosing energy from the gradual emission of GWs.

The reason why these events are of such relevance for cosmology is the possibility of obtaining the value of the luminosity distance for that event directly from the GW, and, from the EM counterpart, its corresponding redshift. With both these measurements it is possible to reconstruct the Hubble diagram, directly constraining the parameters for the model being considered, without relying on a distance ladder.

Although the detection of such events is no shorter than a remarkable achievement, so far there has only been one confirmed SS event, named GW170817 [10], and a proposed EM counterpart to GW190521 [11] was proposed in [12]. Although this first event was measured to incredible accuracy, and showed to be in agreement with ΛCDM, it is not nearly enough to constrain the models we will consider throughout this dissertation. Having in mind forthcoming data from current and future GW observatories, we will develop SS mock catalogs to investigate whether we will be able to distinguish our models from ΛCDM.

## 1.5 Objectives

In short, the main goals of this dissertation are to:

1. Generate realistic mock catalogs of SS events for future and current GW observatories;

2. Understand how cosmological models based on a specific form on non-metricity modify the propagation of GWs;

3. Study two specific cosmological models based on non-metric gravity;

4. See whether future SS events will be able to distinguish between these models and ΛCDM.

Although our analysis is being developed for two specific modified gravity cosmological models, this procedure allows us to forecast the constraints set by SS events for any provided cosmological model.

## 1.6 Structure of this Document

The first three chapters of this dissertation are used to explain the necessary background required to understand the two following chapters, which are the ones that feature original work, while the last chapter includes some final remarks. The outline of each chapter is as follows:

- Chapter 2 will introduce gravity as a geometrical theory, explore some of its objects, and lay the foundations for the non-metric theories of gravity we will consider;

- Chapter 3 will detail the assumptions made regarding the Universe at the cosmological scales, and develop the relevant portions of both ΛCDM and cosmology based on $f(Q)$;

- Chapter 4 introduces the datasets that we have used throughout this dissertation, the procedure employed to generate the SS mock catalogs and the methodology for the Bayesian analysis, for both the model and catalog selection criteria;

- Chapter 5 introduces a model of $f(Q)$ gravity that features a ΛCDM background, which we then forecast using SS events. This chapter led to original work which was published in [13];

- Chapter 6 introduces an $f(Q)$ model with the aim of replacing dark energy. In order to look for viable cosmologies we perform a dynamical system analysis and apply model selection criteria using both SS and SnIa events;

- Chapter 7 includes an outline of this dissertation and presents some foreseeable future work.

# Chapter 2

# Gravity as Geometry

In this chapter, we will introduce and briefly explore different geometrical objects, curvature, torsion and non-metricity, that we use to formulate different theories of gravity, some of which are equivalent to GR.

## 2.1 Geometrical Objects

It is known that a spacetime is endowed with two objects that, in principle, are independent from each other [14]. Those objects are the metric, a rank 2 covariant tensor field which written generically as $g_{\mu\nu}$, that is used in order to define lengths and angles. The other object, referred to as the affine connection and represented symbolically by $\Gamma^{\lambda}{}_{\mu\nu}$, will allow us to speak of parallel transport and to define the covariant derivative.

In order to speak of lengths we introduce the square of an infinitesimal displacement between two points, $ds^2$, which is known as "line element" and is given by

$$ds^2 = g_{\mu\nu}dx^\mu dx^\nu . \tag{2.1}$$

Even though the metric tensor depends on the coordinates of spacetime, we will not make explicit reference to it, leaving the dependency implicit. The metric is used to define the length of a contravariant vector $V^\mu$ as

$$V^2 = g_{\mu\nu}V^\mu V^\nu . \tag{2.2}$$

The metric will also allow us to speak of angles between two contravariant vectors, where the cosine between the vector $V^\mu$ and the vector $U^\nu$ is

$$cos(V,U) = \frac{g_{\mu\nu}V^\mu U^\nu}{\sqrt{g_{\mu\nu}V^\mu V^\nu g_{\rho\sigma}U^\rho U^\sigma}} . \tag{2.3}$$

Finally, for a non-singular metric, that is a metric where its determinant $g \equiv det(g_{\mu\nu})$ is non zero, the inverse of $g_{\mu\nu}$ is given by $g^{\mu\nu}$ such that

$$g_{\alpha\nu}g^{\nu\beta} = \delta^\beta_\alpha . \tag{2.4}$$

The metric and its inverse allow us to raise and lower indices on other quantities, e.g. $V_\mu = g_{\mu\nu}V^\nu$.

Unlike what happens in an Euclidean geometry, in a curved spacetime two nearby points might not necessarily share the same tangent space, which means that it is not possible to define parallel transport in the usual way. We can introduce the affine connection, $\Gamma^\mu{}_{\alpha\beta}$, to relate two nearby tangent spaces by assuming that the displacement is bilinear with both the contravariant vector, which is being propagated, and the displacement. Formally, if we take a contravariant vector $V^a(x)$ and transport it in parallel to $x + \delta x$, $\tilde{V}^a(x + \delta x)$, we can relate the two with

$$V^\mu(x) - \tilde{V}^\mu(x + \delta x) = \Gamma^\mu{}_{\alpha\beta} V^\alpha(x) \delta x^\beta \,. \tag{2.5}$$

Computing the limit when the displacement goes to zero, allow us to define the covariant derivative as

$$\nabla_\mu V^\nu \equiv \partial_\mu V^\nu + \Gamma^\nu{}_{\alpha\mu} V^\alpha \,, \tag{2.6}$$

Where $\partial_\mu \equiv \partial/\partial x^\mu$ is the partial derivative with respect to the coordinate $x^\mu$. If we assume that the covariant derivative has to obey the Leibniz rule, it follows that for a covariant vector

$$\nabla_\mu V_\nu = \partial_\mu V^\nu - \Gamma^\alpha{}_{\nu\mu} V_\alpha \,. \tag{2.7}$$

By contrast to what happens with partial derivatives, if we apply two successive covariant derivatives to a vector, it does not, in general, give the same result as when the order of the derivatives is reversed. That is to say that, in general, the commutator does not vanish, and is given by

$$[\nabla_\mu, \nabla_\nu] V^\alpha = R^\alpha{}_{\beta\mu\nu} V^\beta + T^\sigma{}_{\mu\nu} \nabla_\sigma V^\alpha \,, \tag{2.8}$$

where we introduce the Riemann tensor as

$$R^\sigma{}_{\rho\mu\nu} \equiv \partial_\mu \Gamma^\sigma{}_{\nu\rho} - \partial_\nu \Gamma^\sigma{}_{\mu\rho} + \Gamma^\alpha{}_{\nu\rho} \Gamma^\sigma{}_{\mu\alpha} - \Gamma^\alpha{}_{\mu\rho} \Gamma^\sigma{}_{\nu\alpha} \,. \tag{2.9}$$

and the torsion, which encodes the asymmetric part of the affine connection, defined as

$$T^\lambda{}_{\mu\nu} \equiv \Gamma^\lambda{}_{\mu\nu} - \Gamma^\lambda{}_{\nu\mu} \,. \tag{2.10}$$

Generically, the metric and the affine connection are independent from each other. One can decompose the affine connection as a sum of three different contributions [15]

$$\Gamma^\lambda{}_{\mu\nu} = \left\{^\lambda{}_{\mu\nu}\right\} + K^\lambda{}_{\mu\nu} + L^\lambda{}_{\mu\nu} \,, \tag{2.11}$$

where the first term is the Levi-Civita connection, depending only on the metric and its derivatives, which is defined as

$$\left\{^\lambda{}_{\mu\nu}\right\} \equiv \frac{1}{2} g^{\lambda\beta} \left(\partial_\mu g_{\beta\nu} + \partial_\nu g_{\beta\mu} - \partial_\beta g_{\mu\nu}\right) \,, \tag{2.12}$$

the second term is referred to as the contortion, which is defined as

$$K^\lambda{}_{\mu\nu} \equiv \frac{1}{2} g^{\lambda\beta} \left(T_{\mu\beta\nu} + T_{\nu\beta\mu} + T_{\beta\mu\nu}\right) \,, \tag{2.13}$$

and the third object is referred to as disformation, which is defined by

$$L^\lambda{}_{\mu\nu} \equiv \frac{1}{2}g^{\lambda\beta}\left(-Q_{\mu\beta\nu} - Q_{\nu\beta\mu} + Q_{\beta\mu\nu}\right). \tag{2.14}$$

It is possible to see that the contortion includes all the contributions coming from the torsion tensor, whereas disformation encodes all of the contributions coming from the existence of non-metricity, that is, the covariant derivative of the metric, which is formally defined as

$$Q_{\alpha\mu\nu} \equiv \nabla_\alpha g_{\mu\nu} = \partial_\alpha g_{\mu\nu} - \Gamma^\beta{}_{\alpha\mu}g_{\beta\nu} - \Gamma^\beta{}_{\alpha\nu}g_{\mu\beta}. \tag{2.15}$$

Each of the previously defined geometrical objects, that is, curvature, torsion and non-metricity, have a unique and distinct effect on the propagation of vectors in spacetime. It is in fact possible to provide some geometrical intuition behind each geometrical object, which we represent diagrammatically in fig. 2.1, and summarize as follows [16]:

- Curvature: Measures the rotation experienced by a vector when subject to parallel transport along a closed curve;

- Torsion: Measures the non-closure of the parallelogram when two infinitesimal vectors are subject to parallel transport along each other;

- Non-metricity: Measures how much of the length of the vector changes when subject to parallel transported along any given curve.

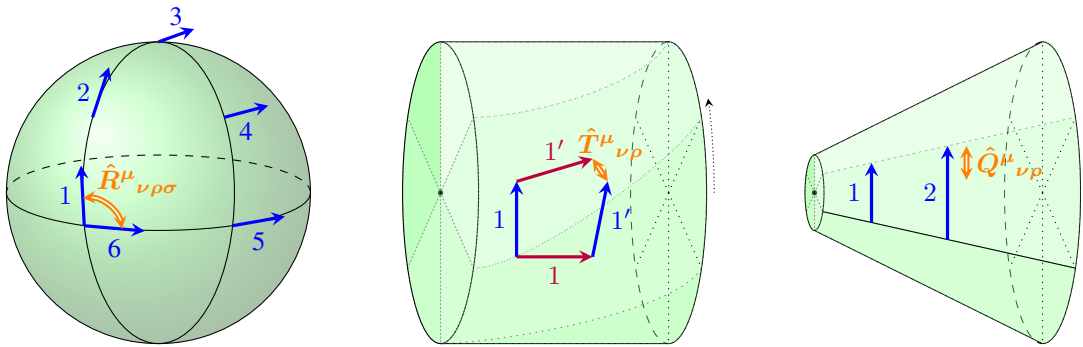

Figure 2.1: Diagrammatic representation of the effect of curvature, torsion and non-metricity on a vector when subject to parallel transport. Taken from [17]

In order to construct a theory of gravity, one is required to choose which geometrical objects are at play in our spacetime. General Relativity (GR) singles out the curvature, presented in eq. (2.9), and assumes $K^\lambda{}_{\mu\nu} = L^\lambda{}_{\mu\nu} = 0$ in eq. (2.11). However, one can choose to work with geometrical objects other than the curvature, allowing us to create theories of gravity that rely on non-metricity or torsion. In fact, we could, in practice, choose two or more objects to fix the underlying geometry, and create theories of gravity which rely on more than one geometrical object. A schematic representation of the different theories of gravity that can be built using the different geometrical objects at stake can be seen in fig. 2.2.

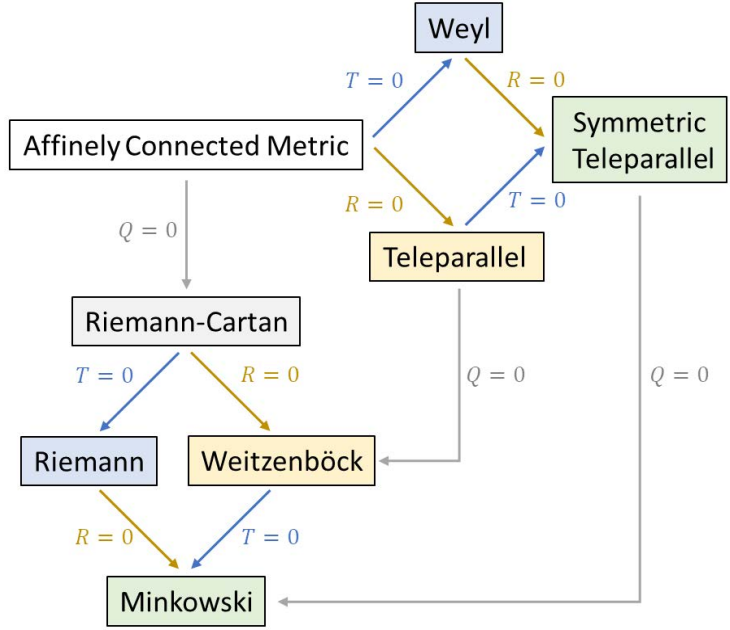

Figure 2.2: Schematic representation of the different possible spacetime geometries built using curvature, torsion and non-metricity. Taken from [18].

## 2.2 The Trinity of Gravity

One of the most remarkable consequences of the interplay between the different geometrical objects, is that it is entirely possible to build geometrical theories of gravity which are equivalent to GR, but without relying on curvature, and instead making use of either non-metricity or torsion.

To provide a glimpse of this equivalence, we will introduce the notation and ideas that were introduced in [19]. When we are considering a given geometrical object, dependent on the connection, keeping null torsion and null non-metricity, we will label it with an "LC" on top of it, to denote it only depends on the contributions coming from the Levi-Civita connection. Similarly, when considering a curvature and non-metricity spacetime, we will use the label "W", to denote a Weitzenböck geometry. For a geometry with no curvature nor torsion we will use the label "ST", which represents the Symmetric Teleparallel setting.

Using the previous notation, one can rewrite the Riemann tensor, presented in eq. (2.9), by separating the contributions coming from the Levi-Civita connection from the contributions due to torsion and non-metricity. The Riemann tensor now reads

$$R^\sigma{}_{\rho\mu\nu} = \overset{\text{LC}}{R}{}^\sigma{}_{\rho\mu\nu} + \overset{\text{LC}}{\nabla}_\mu M^\sigma{}_{\nu\rho} - \overset{\text{LC}}{\nabla}_\nu M^\sigma{}_{\mu\rho} + M^\alpha{}_{\nu\rho} M^\sigma{}_{\mu\alpha} - M^\alpha{}_{\mu\rho} M^\sigma{}_{\nu\alpha}\,, \tag{2.16}$$

where $M^\lambda{}_{\mu\nu}$ includes all contributions coming from both torsion and non-metricity, and is defined as

$$M^\lambda{}_{\mu\nu} \equiv K^\lambda{}_{\mu\nu} + L^\lambda{}_{\mu\nu}\,. \tag{2.17}$$

Additionally, one can define a tensorial quantity which is known as the Ricci tensor

$$R_{\mu\nu} = R^{\alpha}{}_{\mu\alpha\nu}\,, \tag{2.18}$$

that can be used to define the Ricci scalar in the following manner

$$R \equiv g^{\mu\nu} R_{\mu\nu}\,. \tag{2.19}$$

Computing the value of the Ricci scalar explicitly taking into consideration the form of the Riemann tensor presented in eq. (2.16), yields

$$R = \overset{\text{LC}}{R} + M^{\alpha}{}_{\nu\rho} M^{\mu}{}_{\mu\alpha} g^{\nu\rho} - M^{\alpha}{}_{\mu\rho} M^{\mu}{}_{\nu\alpha} g^{\nu\rho} + \overset{\text{LC}}{\nabla}_{\mu} \left( M^{\mu}{}_{\nu\rho} g^{\nu\rho} - M^{\nu}{}_{\nu\rho} g^{\mu\rho} \right)\,. \tag{2.20}$$

If we restrict the geometry of our spacetime to have vanishing torsion and non-metricity, therefore only including contributions due to curvature, we fall back to the setting of GR, where the curvature scalar reduces to the familiar case where

$$R = \overset{\text{LC}}{R}\,. \tag{2.21}$$

In order to have a glimpse of the equivalence, it is useful to remember that it is possible to compute the equations of motion for GR by applying the principle of least action to the so called Einstein-Hilbert action

$$S = \int \sqrt{-g} \left( \frac{c^4}{16\pi G} \overset{\text{LC}}{R} + \mathcal{L}_m \right) dx^4\,, \tag{2.22}$$

where $g$ is the determinant of the metric $g_{\mu\nu}$ and $\mathcal{L}_m$ is the Lagrangian for the energy-matter content of the universe. We can now obtain the field equations for this theory of gravity by performing the variation of the previous action with respect to the metric. Doing so, one obtains what are the well known Einstein Field Equations (EFE)

$$\overset{\text{LC}}{R}_{\mu\nu} - \frac{1}{2} \overset{\text{LC}}{R} g_{\mu\nu} = \frac{8\pi G}{c^4} \mathcal{T}_{\mu\nu}\,, \tag{2.23}$$

where $\mathcal{T}^{\mu\nu}$ is the stress energy tensor

$$\mathcal{T}^{\mu\nu} \equiv \frac{2}{\sqrt{-g}} \frac{\delta(\sqrt{-g}\mathcal{L}_m)}{\partial g_{\mu\nu}}\,. \tag{2.24}$$

If we are to consider a Weitzenböck geometry, meaning that we set both non-metricity and curvature to zero, leaving only torsion to play the role of gravity, eq. (2.20) reduces to

$$\overset{\text{LC}}{R} = -\overset{\text{W}}{T} - 2\overset{\text{LC}}{\nabla}_{\alpha} \overset{\text{W}}{T}{}^{\alpha}\,, \tag{2.25}$$

where $T$ is the torsion scalar, which is defined in a connection independent manner as

$$T \equiv \frac{1}{4} T_{\alpha\beta\gamma} T^{\alpha\beta\gamma} + \frac{1}{2} T_{\alpha\beta\gamma} T^{\gamma\beta\alpha} - T_{\alpha} T^{\alpha}\,, \tag{2.26}$$

and

$$T_{\mu} \equiv T^{\alpha}{}_{\mu\alpha}\,. \tag{2.27}$$

By looking at eq. (2.25), we can see that the curvature scalar for GR is replaced by the torsion scalar in a Weitzenböck geometry plus a divergence term, being acted upon by a covariant derivative. This additional divergence term will be integrated and computed at infinity, i.e. at the boundary, meaning that it does not contribute to the field equations, implying that both of these geometrical settings are equivalent to each other. This geometrical theory of gravity which relies solely on torsion instead of curvature is equivalent to GR, and is named the Teleparallel Equivalent of General Relativity (TEGR).

Similarly, if we set both the curvature and torsion to zero, leaving non-metricity to mediate the gravitational interaction, the curvature scalar presented in eq. (2.20) simplifies to

$$\overset{\text{LC}}{R} = \overset{\text{ST}}{Q} - \overset{\text{LC}}{\nabla}_\alpha (\overset{\text{ST}}{Q}{}^\alpha - \overset{\text{ST}}{\tilde{Q}}{}^\alpha) \,, \tag{2.28}$$

where $Q$ is the non-metricity scalar that is defined as

$$Q \equiv -\frac{1}{4} Q_{\alpha\beta\gamma} Q^{\alpha\beta\gamma} + \frac{1}{2} Q_{\alpha\beta\gamma} Q^{\gamma\beta\alpha} + \frac{1}{4} Q_\alpha Q^\alpha - \frac{1}{2} Q_\alpha \tilde{Q}^\alpha \,, \tag{2.29}$$

and the two independent contractions of the non-metricity tensor are

$$Q_\mu \equiv Q_\mu{}^\alpha{}_\mu \,, \qquad \tilde{Q}^\mu \equiv Q_\alpha{}^{\alpha\mu} \,. \tag{2.30}$$

By the same arguments that we have used when we had torsion instead of curvature, we can see that the curvature scalar in eq. (2.28) differs only from the non-metricity scalar by a boundary term, implying that this theory is equivalent to GR. This non-metric theory of gravity is referred to as the Symmetric Teleparallel Equivalent of General Relativity (STEGR).

The same procedure we have applied in order to obtain the EFE can be applied for the case of STEGR, where the action will have the non-metricity scalar computed for a flat and torsion-free spacetime, $\overset{\text{ST}}{Q}$, instead of the curvature scalar computed by relying only on the Levi-Civita connection $\overset{\text{LC}}{R}$. Naturally, this also applies for TEGR, where the torsion scalar in a metric and flat spacetime $\overset{\text{W}}{T}$ is used instead of $\overset{\text{LC}}{R}$.

The equivalence between these three different theories of gravity, or has it is often referred to as, the three different interpretations of gravity, have been recently popularized as the geometrical trinity of gravity [16].

## 2.3 Generalization of the Trinity of Gravity

As we have briefly mentioned in the introduction, the geometrical trinity of gravity is a framework which one can use as a starting point to create modified theories of gravity, knowing that the advancements developed by GR will be present.

Although there is a countless number of ways to generalize any given theory, perhaps the most simple way to do so is to promote the scalar quantity in the action for each of the different representations of gravity, to an arbitrary function of itself.

We can apply this procedure to generalize the Einstein-Hilbert action, which leads to the well known and widely studied class of $f(R)$ theories (for a review see e.g.: [20]). Likewise, one can do the same for a spacetime without curvature and non-metricity, generalizing the TEGR to a particular case of Teleparallel Gravity (TG), which we refer to $f(T)$ gravity. This class of theories have also been widely studied, and for a review on the gravity and cosmology behind $f(T)$ we refer the reader to [17]. Of interest to us in this work will be the generalization of STEGR, leading us to $f(Q)$ gravity, a particular case of a more general class of theories which are referred to as the Symmetric Teleparallel Gravity (STG). The action for this theory of gravity reads

$$S = \int \sqrt{-g} \left[ -\frac{c^4}{16\pi G} f(Q) + \mathcal{L}_m \right] d^4x \,. \tag{2.31}$$

When varying the previous action with respect to the metric, one obtains the field equations [7]

$$\frac{2}{\sqrt{-g}} \nabla_\alpha \left( \sqrt{-g} f_Q(Q) P^{\alpha\mu}{}_\nu \right) + \frac{1}{2} \delta^\mu_\nu f + f_Q(Q) P^{\mu\alpha\beta} Q_{\nu\alpha\beta} = \frac{8\pi G}{c^4} \mathcal{T}^\mu{}_\nu \,, \tag{2.32}$$

where $f_Q(Q) \equiv \partial f(Q)/\partial Q$, $L^\alpha{}_{\mu\nu}$ is the disformation tensor which we have already met previously and $P^{\alpha\mu\nu}$ is the non-metricity conjugate

$$P^\alpha{}_{\mu\nu} = -\frac{1}{2} L^\alpha{}_{\mu\nu} + \frac{1}{4}(Q^\alpha - \tilde{Q}^\alpha) - \frac{1}{4} \delta^\alpha_{(\mu} Q_{\nu)} \,, \tag{2.33}$$

What is surprising from these generalizations is that, unlike what we expected given the arbitrary nature of the functions $f(Q)$, $f(T)$ and $f(R)$, it is still possible to show that there is a set of assumptions one can make such that the field equations for these theories are the same. In [21] the authors show that under the assumption of a flat, homogeneous and isotropic universe, which is permeated by a perfect fluid, then the field equations derived for $f(Q)$ and $f(T)$ cosmology are formally equivalent, as well as the relationship between the scalar quantity in the action and the Hubble function. Additionally, by decomposing the curvature scalar into the sum of two terms $R = G + B$, where $G$ is a bulk term and $B$ a boundary term, it is also shown that $f(G)$ cosmology is also formally equivalent to the two previous theories. In the spirit of the previous equivalence between geometrical theories this equivalence is, in a sense, a "cosmological trinity of gravity". This is of interest to us because, as we will see further ahead in chapter 3, those are the baseline assumptions for the work developed throughout this dissertation.

An important remark we wish to make is that, due to the lack of coupling between the gravitational sector and the matter content of the universe, these three generalized theories of gravity do not modify the fact that the covariant derivative of the energy-momentum tensor is equal to zero, as shown in [22].

For the scope of this dissertation, only models based on non-metricity will be considered. To simplify this notation we will often refer to the function $f(Q)$ simplify as $f$ and its derivative as $f_Q$. We will, however, still include explicit dependence on the non-metricity scalar to denote gravitational models based solely on non-metricity, which have an action of the form of the one presented in eq. (2.31). Additionally, because curvature and torsion are always set to zero, the label "ST" is always assumed to be there on matters of geometry, although no explicit reference to it will be made.

# Chapter 3

# Cosmological Setting

In this chapter, we detail the assumptions made regarding the universe at cosmological scales, which we then use to develop the relevant topics of both the standard model of cosmology, ΛCDM, and a cosmological model based on $f(Q)$ gravity, keeping the function $f(Q)$ to be completely generic. Unless otherwise stated, the contents in this chapter will be based on [23] and [24].

## 3.1 The Metric at Large Scales

One of the most fundamental pieces of our knowledge regarding the universe at the large scales, is the assumption that it is homogeneous, i.e. no point in space is privileged when compared to other, and isotropic, i.e. from a given point in space the Universe looks the same regardless of the direction we are looking at. This assumption is known as the cosmological principle, a well observationally motivated assumption of the distribution of matter in the universe at cosmological scales. Under these assumptions, the most general line element is given by

$$ds^2 = -c^2 dt^2 + a^2(t)\left(\frac{dr^2}{1-kr^2} + r^2(d\theta^2 + \sin^2\theta d\phi^2)\right), \tag{3.1}$$

which is know as the Friedmann Lemaître Robertson Walker (FLRW) metric, where $a(t)$ is the scale factor, which is a function of the cosmic time $t$, and $k \in (-1, 0, 1)$ is the spatial the curvature of the universe.

One can simplify this metric even further considering a flat universe, something which is supported by the Planck collaboration in [25]. The line element then reduces to

$$ds^2 = -c^2 dt^2 + a^2(t)(dx^2 + dy^2 + dz^2). \tag{3.2}$$

The scale factor, which describes an expanding or contracting universe, might cause the frequency of photons emitted in the past, $\nu_e$, to be different to the frequency that we measure today, $\nu_0$. For a photon emitted in the past at time $t_e$, and is measured today at time $t_0$, its frequency changes such that

$$\frac{\nu_e}{\nu_0} = \frac{a_0}{a_e}, \tag{3.3}$$

A useful quantity to state by how much has the universe grown is the redshift, that tells us how much the frequency for a given photon as decreased, and is defined as

$$1 + z \equiv \frac{\nu_e}{\nu_0}, \quad (3.4)$$

which when considered together with the result obtained in eq. (3.3), reveals that

$$1 + z = \frac{a_0}{a_e}. \quad (3.5)$$

As convection dictates, we consider the scale factor today to be $a_0 = 1$. This can be done without loss of generality, since we can always redefine the spatial coordinates.

## 3.2 Cosmological Fluid

One can show that the requirements set by the cosmological principle, regarding homogeneity and isotropy, can be satisfied by a fluid with no shear stress, nor viscosity, meaning that it is fully characterized by its energy density and pressure. The energy-momentum tensor of a perfect fluid is

$$\mathcal{T}_{\mu\nu} = (\rho c^2 + P) u_\mu u_\nu + P g_{\mu\nu}, \quad (3.6)$$

where $\rho$ and $P$ are the total fluid energy density and pressure respectively, while $u_\mu$ is the fluid four-velocity relative to the observer. For a comoving observer the four-velocity is given by $u^\mu = (-1, 0, 0, 0)$, which simplifies the energy-momentum tensor to be strictly diagonal, reducing to

$$\mathcal{T}^\mu{}_\nu = -(\rho c^2 + P)\delta^\mu_0 \delta^0_\nu + P \delta^\mu_\nu. \quad (3.7)$$

In this dissertation, we consider the existence of three different components in the perfect fluid:

- Radiation: This includes contributions from all relativistic species where the pressure can be related to the energy density by $P = \rho c^2 / 3$;

- Matter: All the contents in the universe which exhibit a negligible pressure, $P \approx 0$;

- Dark Energy: An exotic component which has a constant energy density value throughout the history of the universe and that exhibits negative pressure of the form $P = -\rho c^2$.

We can now parametrize each component of the perfect fluid with an equation of state (EoS) of the form

$$w \equiv \frac{P}{\rho c^2}, \quad (3.8)$$

which after computing the value of $w$ reveals that

$$w = \begin{cases} 1/3, & \text{radiation} \\ 0, & \text{matter} \\ -1, & \text{dark energy} \end{cases} \quad (3.9)$$

which we will use further ahead to compute the evolution of the energy density for each component of the perfect fluid.

## 3.3 The Standard Cosmological Model

The standard model of cosmology, from now on referred to as ΛCDM, is the dominant paradigm in modern cosmology, being able to successfully account for most of the features of the universe observed at cosmological scales. Its main features are [5]

1. The universe consists of matter, radiation and dark energy;

2. GR is the theory to describe gravity at the cosmological scales;

3. The cosmological principle is regarded as true;

4. The universe is considered to be spatially flat;

5. There is a period of initial inflation in the primordial universe;

6. There are 6 independent parameters: The baryon ($\Omega_b$) and cold dark matter ($\Omega_c$) densities, the optical depth at reionization $\tau$, the amplitude $A_s$ and tilt $n_s$ of the primordial scale fluctuations and the Hubble constant $h$.

In the previous paragraph, as well as throughout the rest of this work, $h$ refers to the dimensionless Hubble constant, defined such that $H_0 = 100 h \, \text{km}\,\text{s}^{-1}\text{Mpc}^{-1}$.

By applying the FLRW metric, presented in eq. (3.2), the energy-momentum tensor of the perfect fluid, given in eq. (3.7), and inserting both in the EFE, presented in eq. (2.23), one is able to derive what are known as the Friedmann equations. The time-time component of the result arising from this computation reveals what is referred to as the first Friedmann equation, which is given by

$$H^2 = \frac{8\pi G}{3}\rho \,, \tag{3.10}$$

while the second Friedmann equation, also known as the Raychaudhuri equation or acceleration equation, is obtained from the spatial components of the previous computation, given by

$$\frac{\ddot{a}}{a} = -\frac{4\pi G}{3}\left(\rho + 3\frac{P}{c^2}\right) \,. \tag{3.11}$$

From the covariant derivative of the energy-momentum tensor, we obtain what is known as the continuity equation

$$\dot{\rho} + 3\frac{\dot{a}}{a}\left(\rho + \frac{P}{c^2}\right) = 0 \,, \tag{3.12}$$

where the overdot indicates a derivative with respect to the cosmic time.

By the means of the continuity equation, we are able to compute the evolution of each of the components of the cosmological fluid, with respect to the scale factor. This is done by taking the EoS, $P = w\rho$, and inserting it in the continuity equation, reducing to

$$\frac{\dot{\rho}}{\rho} = -3\frac{\dot{a}}{a}(1+w) \,. \tag{3.13}$$

By assuming that the value of $w$ is a constant, which is true for all components of the perfect fluid, we can integrate the previous equation and show that for the $i$-th component the evolution of the energy density is given by

$$\rho_i = \rho_{i,0}\, a^{-3(1+w)} \,, \tag{3.14}$$

where the index 0 is once again used to refer to the value of that quantity at the present time.

Using the previous result, and the value of $w$ previously computed in eq. (3.9), we can show that the density evolution for radiation, which from now on we will denote with an index $\gamma$, is

$$\rho_\gamma = \rho_{\gamma,0}\, a^{-4} \,. \tag{3.15}$$

Similarly, the density evolution for matter, denoted by the index $m$, is

$$\rho_m = \rho_{m,0}\, a^{-3} \,, \tag{3.16}$$

and finally, the density evolution of dark energy, denoted using the index $\Lambda$, is

$$\rho_\Lambda = \rho_{\Lambda,0} \,. \tag{3.17}$$

We now define the relative abundance of the $i$ component of the cosmological fluid for the present day, which will be used from now on instead of the individual energy densities, represented by $\Omega_i$, as

$$\Omega_i \equiv \frac{8\pi G}{3H_0^2}\rho_{i,0} \,, \tag{3.18}$$

where $H \equiv \dot{a}/a$ is the Hubble function and $H_0 \equiv H(a=a_0=1)$ is the Hubble constant.

## 3.4 Cosmology in $f(Q)$

In this section we will drop $\Lambda$CDM second postulate and instead of working with GR we will consider instead an arbitrary $f(Q)$ modified gravity cosmological model.

Computing the non-metricity scalar, presented in eq. (2.29), for the metric in eq. (3.2), we have that [7]

$$Q = 6H^2 \,. \tag{3.19}$$

Similarly to what was done before, we now insert the FLRW metric and the energy-momentum tensor of the perfect fluid in the modified field equations for $f(Q)$, presented in eq. (2.32). We obtain a modified version of the first Friedmann equation, which reads [26]

$$6f_Q H^2 - \frac{1}{2}f = 8\pi G\rho \,, \tag{3.20}$$

and the modified second Friedmann equation is

$$(12H^2 f_{QQ} + f_Q)\dot{H} = -4\pi G \left(\rho + \frac{P}{c^2}\right), \quad (3.21)$$

where the index $Q$ denotes a partial derivative with respect to $Q$.

To understand the dynamics of a universe in $f(Q)$ gravity, we must also understand how each of our cosmological fluid behaves. Given that for $f(Q)$ gravity the covariant derivative of the energy-momentum tensor is still valid, regardless of the form that the function might take, then it immediately follows that the continuity equation

$$\dot{\rho} + 3\frac{\dot{a}}{a}\left(\rho + \frac{P}{c^2}\right) = 0, \quad (3.22)$$

still holds true. As such, this implies that the cosmological fluids behave precisely in the same way as in $\Lambda$CDM.

## 3.5 Propagation of Gravitational Waves

One of the most remarkable predictions of GR, which is extended to the other geometrical theories of gravity, is the existence of gravitational waves (GWs). These waves are tensorial perturbations on top of the background of spacetime which propagate outwards from their source, interacting with the energy-matter content of the universe along the way. At present time the main source of GW events is the merger of binary systems of compact bodies, which are known to be unstable due to the gradual emission of GWs. Although their existence should be verified across different theories of gravity, given that there is more than enough empirical evidence supporting their existence, the way that they propagate through spacetime might not necessarily be the same between the different theories. Given that we are interested in using standard siren (SS) events as a probe to understand the evolution of the cosmos, here we will see just how much does $\Lambda$CDM and $f(Q)$ based cosmology differ in the propagation of GWs.

For the scope of this dissertation, we are interested in knowing how does a GW emitted from a SS event reach the Earth. Knowing that the events we expect to observe, the merger of compact binary systems, are located very far away, then we can safely work under the assumptions that GWs are small tensorial perturbations on top of an otherwise locally flat Minkowski spacetime. Under these assumptions we can decompose the metric tensor as [27]

$$g_{\mu\nu} = \eta_{\mu\nu} + h_{\mu\nu}, \quad (3.23)$$

where $\eta_{\mu\nu}$ is the Minkowski metric and $h_{\mu\nu}$ is a small perturbation such that $|h_{\mu\nu}| \ll 1$.

From the linearized EFE it is possible to see that, under the proper gauge, a solution for $h_{\mu\nu}$ can take the shape of a plane wave. Without loss of generality we take our wave to propagate along the $z$ axis and the solution for $h_{\mu\nu}$ tensor takes the form

$$h_{\mu\nu} = \begin{bmatrix} 0 & 0 & 0 & 0 \\ 0 & h_+ & h_\times & 0 \\ 0 & h_\times & -h_+ & 0 \\ 0 & 0 & 0 & 0 \end{bmatrix} \cos\left(kc(t - z/c) + \phi_0\right), \quad (3.24)$$

where $k$ is the modulus of the wave vector of the GW and $\phi_0$ is an arbitrary phase.

The reason why the two degrees of freedom for this metric are labeled as $h_+$ and $h_\times$, often referred to as polarizations, is due to the way they compress and expand a circle of point-like test particles. The $h_+$ polarization modifies the circle in what resembles a plus sign, with one direction compressing and the other expanding, whereas $h_\times$ does the same but in a cross like manner.

The propagation of tensorial perturbations in a cosmological background for a $\Lambda$CDM universe is given by [28]

$$\bar{h}_A'' + 2\mathcal{H}\bar{h}_A' + k^2 \bar{h}_A = 0 \,, \tag{3.25}$$

where $\bar{h}_A$ are the Fourier modes of the GW amplitude, $A = +, \times$ represent the polarization, the prime denotes a derivative with respect to conformal time $\eta$, which is defined as $d\eta = dt/a$, and $\mathcal{H} = a'/a$.

Just like electromagnetic (EM) radiation is redshifted when propagating throughout the universe, as a consequence of the expansion, gravitational radiation, i.e. GW, suffers the same effect. This means that it looses energy as it travels in an expanding universe and, similarly to the EM radiation, it suffers from a redshift effect. The luminosity distance for a GW under this condition is similar to the luminosity distance of an EM event, which reads

$$d_L(z) = (1+z)c \int_0^z \frac{1}{H(z)} dz \,. \tag{3.26}$$

However, eq. (3.25) does not hold when we replace GR by a model of $f(Q)$ gravity. Under the same assumptions that we have used to obtain the previous equation, but replacing GR by an arbitrary $f(Q)$ theory of gravity, the propagation of the tensorial perturbations is [7]

$$\bar{h}_A'' + 2\mathcal{H}(1 + \delta(z))\bar{h}_A' + k^2 \bar{h}_A = 0 \,, \tag{3.27}$$

which differs from the case of $\Lambda$CDM by a friction term, $\delta(z)$, that for an $f(Q)$ cosmological model takes the form

$$\delta(z) = \frac{d \ln f_Q}{2\mathcal{H} d\eta} \,. \tag{3.28}$$

Following [29], in order to eliminate the friction term and obtain an expression that is formally equivalent to eq. (3.25), we introduce a modified scale factor $\tilde{a}$ that obeys the equation

$$\frac{\tilde{a}'}{a} = \mathcal{H}(1 + \delta(z)) \,. \tag{3.29}$$

By integrating both sides we obtain that the ratio between the modified and the usual scale factor is

$$\frac{\tilde{a}}{a} = \exp\left(\int_0^z \frac{\delta(z)}{1+z} dz\right) \,. \tag{3.30}$$

Given that in an FLRW metric the value of $h_A$ is inversely proportional to the scale factor, and we obtain an expression that is formally equivalent in $f(Q)$ cosmology only with a modified scale factor, this leads to a modification to the luminosity distance of the GW that takes the form

$$d_L^{(\text{GW})}(z) = \exp\left(\int_0^z \frac{\delta(z)}{1+z} dz\right) d_L(z), \tag{3.31}$$

where $d_L(z)$ is the same EM luminosity distance we saw in eq. (3.26). Inserting the value of $\delta$ for an $f(Q)$ cosmological model, we can now write

$$d_L^{(\text{GW})}(z) = \sqrt{\frac{f_Q^{(0)}}{f_Q}}\, d_L(z), \tag{3.32}$$

where $f_Q^{(0)}$ is the function $f_Q$ computed at the present day.

# Chapter 4

# Datasets

In this chapter we introduce the datasets used throughout this dissertation, the Bayesian inference methodology as well as the model and catalog selection criteria.

## 4.1 Type Ia Supernovae

A type Ia supernova (SnIa) is an event that takes place when a white dwarf in a binary system accretes mass from its companion beyond the Chandrasehkar limit. The result is an explosion of considerable and predictable brightness, which is of great use to study the Universe at cosmological scales.

The usage of SnIa on this dissertation will be to provide an additional source of data to complement the constrains set by SS events. There are two reasons why we are forced to add an additional source of data. The first is because the model which we will study in the chapter 5 features a degeneracy between two of its parameters in the expression for the luminosity distance for GWs, requiring an additional source of data to fix one of its parameters. The second reason is to assist in the model selection which will be carried out in chapter 6, between the model which is being studied in that chapter and $\Lambda$CDM.

The relationship between the apparent magnitude and the luminosity distance is given by

$$m = M + 5\log\left(d_L(z)\right) + 25\,, \tag{4.1}$$

where $M$ is the bolometric magnitude.

Based on [30], we start by taking a standard $\chi^2$ of the form

$$\chi^2 = \sum_{i=1}^{N}\left[\frac{m^{(\text{obs})}(z_i) - m(z_i)}{\sigma(z_i)}\right]^2\,, \tag{4.2}$$

where $m^{(\text{obs})}$ is the observed apparent magnitude, $m$ is the theoretical prediction for the apparent magnitude, $N$ the total number of SnIa events and finally $z_i$ and $\sigma(z_i)$ are both the redshift and the corresponding error for the $i$-th event.

Equation (4.1) can then be rewritten as

$$m = \mathcal{M} + 5\log\left(D_L(z)\right)\,, \tag{4.3}$$

where

$$D_L(z) \equiv \frac{H_0}{c} d_L(z) = (1+z) \int_0^z \frac{1}{E(z)} dz, \tag{4.4}$$

is referred to as the $H_0$ independent luminosity distance and $\mathcal{M}$ is defined as

$$\mathcal{M} \equiv 25 + M + 5 \log\left(\frac{c}{H_0}\right). \tag{4.5}$$

We can see that the parameter $M$ is degenerate with $H_0$, we cannot fit either of these parameters separately. Since $\mathcal{M}$ is a nuisance parameter we marginalize it out of the likelihood.

By inserting the apparent magnitude written as presented in eq. (4.3) in the expression for the $\chi^2$ we obtain that

$$\chi^2 = \sum_{i=1}^{N} \left[ \frac{\Delta^2(z_i)}{\sigma^2(z_i)} - 2\frac{\Delta(z_i)\mathcal{M}}{\sigma^2(z_i)} + \frac{\mathcal{M}^2}{\sigma^2(z_i)} \right], \tag{4.6}$$

where we have defined $\Delta(z)$ to be given by

$$\Delta(z) \equiv m^{(\text{obs})} - 5 \log D_L(z). \tag{4.7}$$

which can be further simplified to

$$\chi^2 = A - 2B\mathcal{M} + \mathcal{M}^2 C, \tag{4.8}$$

where $A$, $B$ and $C$ are defined as

$$A \equiv \sum_{i=1}^{N} \frac{\Delta^2(z_i)}{\sigma^2(z_i)}, \qquad B \equiv \sum_{i=1}^{N} \frac{\Delta(z_i)}{\sigma^2(z_i)}, \qquad C \equiv \sum_{i=1}^{N} \frac{1}{\sigma^2(z_i)}. \tag{4.9}$$

We now take the value of $\chi^2$, obtained in eq. (4.8), and considering a Gaussian likelihood we marginalize all contributions coming from the term $\mathcal{M}$. Mathematically this is given by

$$\mathcal{L} = \int_{-\infty}^{\infty} e^{-\chi^2/2} d\mathcal{M}, \tag{4.10}$$

which after integrating reduces to

$$\mathcal{L} = \sqrt{\frac{2\pi}{C}} \; e^{\frac{1}{2}(-A + B^2/C)}. \tag{4.11}$$

Looking at the expression for the likelihood, we note that there is a multiplicative factor which does not depend on the value of the parameters, but only on the value of the sum of the $\sigma_i$. Given that $\sigma_i$ does not depend on the parameters of our model we can safely drop the multiplicative term to the left of the exponential, simplifying the likelihood for the SnIa to

$$\mathcal{L} = e^{\frac{1}{2}(-A + B^2/C)}. \tag{4.12}$$

As for the source of the data, we will be considering SnIa events from the Pantheon sample, as was presented in [31] and is available in a public repository at [32]. For performance reasons, the binned sample was used throughout this dissertation. We verified that this choice of a reduced sample will not affect our results when compared to the complete dataset.

## 4.2 Standard Sirens

A SS event is any astrophysical phenomena that emits radiation both in the GW spectrum, which can be used to directly obtain the luminosity distance of the source, and in the EM spectrum, allowing us to obtain the redshift where the event took place. The main source for SS events are the merger of massive binary systems, something which is bound to happen for every binary system as they are known to be unstable in GR, due to the emission of gravitational waves. By making use of these events, it becomes possible to reconstruct the luminosity distance as a function of redshift, without requiring a ladder distance calibration, making these events prime candidates for testing the evolution of the universe at the very large scales.

It is the aim of this dissertation, we will study two well known gravitational wave observatories: the advanced Laser Interferometer Gravitational-Wave Observatory (LIGO), introduced in [33], which is a set of two second generation ground based gravitational wave observatories, that together with the advanced Virgo, introduced in [34], which is also a second generation ground based observatory, provide the most comprehensive catalog of gravitational wave events to date.

As for future observatories, we will consider the Laser Interferometer Space Antenna (LISA), which aims to be the first GW observatory in space by placing three satellites in solar orbit. It has been recently proposed in [35], with a plan to be launched as early as 2030 and a mission lifetime of 4 years, with a possible extension of up to 10 years. According to [36] this observatory is expected to be able to detect GW events, mostly coming from massive black hole binaries (MBHBs), with redshifts up to $z \approx 10$. The data obtained will translate into a major leap in the evaluation of theoretical cosmological settings, given that current measurements have only measured GW events with a maximum redshift of $z \approx 1$.

Additionally, we will also study the Einstein Telescope (ET), which is a third generation underground gravitational wave observatory with a successful proposal [37], that is aimed to have its first light in 2035. Its focus will be mostly targeted towards looking for new physics in high energy events, namely, by looking at the merger of binary neutron stars (BNSs), with the expectations of having a glimpse of its internal structure. Although these events will be probed at low redshifts, they also provide valuable cosmological insight.

So far only one SS event has been detected so far, named GW170817 [10], and the possible detection of an EM counterpart of GW109521 [11] has been suggested in [12]. Although they were remarkable achievements, with the first event being able to put very tight bounds on the speed of propagation of GWs, neither of them is able to put meaningful constrains on either of our models. As such, we choose to create SS mock catalogs, which will allow us instead to forecast the constrains set on any given cosmological model, for the LIGO-Virgo collaboration, LISA and ET.

The procedure used to generate the SS events is generic, and can be summarized by per-forming the following steps:

1. Obtain a redshift for the event, $z_*$, by sampling the event probability distribution function;

2. Generate the luminosity distance for the obtained redshift, $d_L(z_*)$, using a fiducial cosmological model.

3. Compute the error for the obtained redshift for the corresponding observatory, $\sigma_{\text{tot}}(z_*)$, and consider it to be the $1\sigma$ region for the luminosity distance;

4. Consider the observed value of the luminosity distance, $d_L^{(\text{obs})}(z_*)$, to be a sample from a Gaussian distribution with mean in $d_L(z_*)$ and with standard deviation equal to $\sigma_{\text{tot}}(z_*)$;

5. Repeat this procedure until we achieve the number of events we expect to observe for each of the observatories considered.

The last step is performed to obtain a more realistic catalog, such that the most likely value for the observed luminosity distance does not fall consistently on top of the fiducial value. As for the fiducial cosmology, we have decided to consider ΛCDM with fiducial values $h = 0.7$ and $\Omega_m = 0.284$. This latter value corresponds to the best fit for the SnIa from the Pantheon sample, such that we avoid tensions between SS and SnIa.

Following a Bayesian analysis, we take the likelihood for this dataset to be a Gaussian function, which is written as

$$\mathcal{L} = \prod_{i=1}^{N} \frac{1}{\sqrt{2\pi}\sigma_{\text{tot}}(z_i)} \exp\left(-\frac{1}{2}\left[\frac{d_{\text{GW}}^{(\text{obs})}(z_i) - d_{\text{GW}}(z_i)}{\sigma_{\text{tot}}(z_i)}\right]^2\right), \quad (4.13)$$

where $d_{\text{GW}}^{(\text{obs})}(z)$ is the observed luminosity distance, $d_{\text{GW}}(z)$ is the theoretical gravitational wave luminosity distance predicted by our model and $N$ is the number of SSs events.

### 4.2.1 LIGO-Virgo Forecasts

To obtain the expected probability distribution function of SSs events for the LIGO-Virgo observatories we take the luminosity distance distribution function found in [38], which is then sampled to obtain the corresponding redshift, using the fiducial cosmology. We then follow the steps 4, 5 and 6 of the mock catalog generation procedure which we have outline before.

Following the steps developed in [36], we consider that each LIGO-Virgo catalog is composed of $N = 50$ events and the error as a function of redshift is given by

$$\sigma_{\text{LIGO-Virgo}}^2 = \sigma_{d_L}^2 + \left(\frac{d}{dz}(d_L)\sigma_{\text{spect}}\right)^2, \quad (4.14)$$

where $\sigma_{d_L}$ is a rough approximation based on the lowest signal-to-noise ratio the LIGO-Virgo is expected to measure, given by

$$\sigma_{d_L} = \frac{5.63 \times 10^{-4}}{\text{Mpc}} d_L^2(z), \quad (4.15)$$

and the second contribution is the propagation to the luminosity distance of the error in the redshift measurement, assumed to be of spectroscopic origin,

$$\sigma_{\text{spect}} = 0.005(1+z)\,. \tag{4.16}$$

### 4.2.2 LISA

In [39] the authors presented the redshift distribution of SSs events, which are expected to be visible to LISA. The formation process for MBHB events is still not yet fully understood, therefore the authors considered three distinct populations of MBHBs: *No Delay*, *Delay* and *Pop III*. For a complete description of each of the populations we refer the reader to [40]. The previous redshift distributions were developed for three different mission specifications, throughout this dissertation we have chosen to work with the mission specification L6A2M5N2, since it is the closest to the proposed mission specification presented in [35].

In order to match the analysis developed in [41], we have modified the previous redshift distribution to include no events below $z = 0.1$, justified by the absence of MBHB events in this redshift range. For the sake of convenience, we have fitted each of the redshift distribution functions with a beta distribution, which takes the form

$$f(z) = \gamma \left(\frac{z}{9}\right)^{\alpha-1} \left(1 - \frac{z}{9}\right)^{\beta-1}, \tag{4.17}$$

where the best fit values for each population are shown in table 4.1 and the normalized redshift probability distribution function, as well as the best fit, are represented graphically in fig. 4.1.

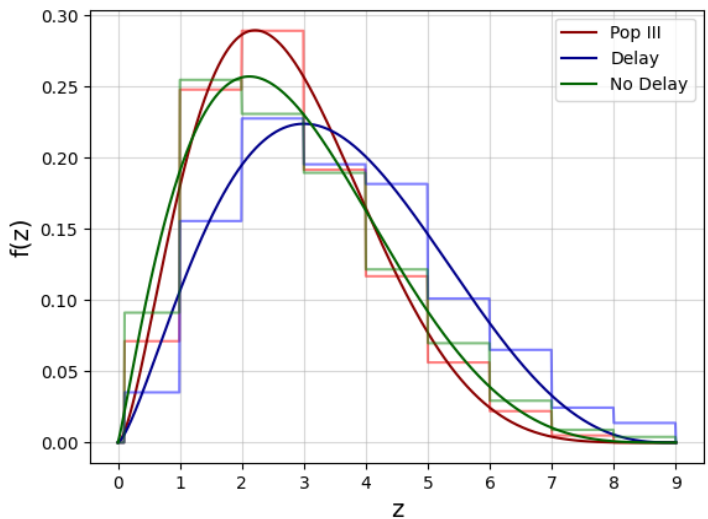

Figure 4.1: Expected normalized redshift distribution for LISA SS, for the L6A2M5N2 mission specification, for populations *Pop III*, *Delay* and *No Delay*, fitted with eq. (4.17), with the best fit values present in table 4.1.

Based on [41], we consider that the total error for the luminosity distance as a function of redshift for LISA is given by

$$\sigma_{\text{LISA}}^2 = \sigma_{\text{delens}}^2 + \sigma_{\text{v}}^2 + \sigma_{\text{inst}}^2 + \left(\frac{d}{dz}(d_L)\sigma_{\text{photo}}\right)^2, \tag{4.18}$$

where the first term corresponds to the total lensing contribution, which can be decomposed in the multiplication of two factors, given by

$$\sigma_{\text{delens}} = F_{\text{delens}}\,\sigma_{\text{lens}}, \tag{4.19}$$

where

$$\sigma_{\text{lens}} = 0.066 \left(\frac{1 - (1+z)^{-0.25}}{0.25}\right)^{1.8} d_L(z), \tag{4.20}$$

is the analytically estimated weak lensing contribution and

$$F_{\text{delens}} = 1 - \frac{0.3}{\pi/2}\arctan\left(z/0.073\right), \tag{4.21}$$

is the delensing factor, which includes the possibility of estimating the lensing magnification distribution and partially correct the effect that weak lensing produces on the measurement of the GWs.

|          | $\alpha$ | $\beta$ | $\gamma$ |
|----------|------|------|-------|
| Pop III  | 2.64 | 6.03 | 11.95 |
| Delay    | 2.42 | 3.84 | 3.37  |
| No Delay | 2.14 | 4.7  | 3.61  |

Table 4.1: Best fit values of the beta distribution, given by eq. (4.17), for the redshift distribution of the MBHB populations *Pop III*, *Delay* and *No Delay*, presented graphically in figure 4.1.

The second term considers the error which originates due to the peculiar velocity of the sources, and is estimated to be given by

$$\sigma_{\text{v}} = \left[1 + \frac{c(1+z)^2}{H(z)d_L(z)}\right]\frac{500\text{ km/s}}{c}d_L(z). \tag{4.22}$$

The third term of the error includes the LISA instrumental error on the measurement of the luminosity distance, which can be estimated by

$$\sigma_{\text{inst}} = 0.05\left(\frac{d_L^2(z)}{36.6\text{Gpc}}\right), \tag{4.23}$$

and finally, the last term, includes the redshift error associated with photometric measurements, which is then propagated to the luminosity distance given the fiducial cosmology. We assume this source of error to only take place at redshifts larger than 2, and takes the form

$$\sigma_{\text{photo}} = 0.03(1+z),\text{ if } z > 2. \tag{4.24}$$

As for the population, according to [41], the major difference is expected to come from the number of events detected. As such, we have decided to work with the *No Delay* population, as it seems to provide a middle ground between the other two with respect to the redshift distribution of events.

We then take the most conservative estimate of events expected to detect, for all the MBHB populations, which, according to [42], when using the current hardware specification and a 4 year mission lifetime is $N = 15$ SS events.

### 4.2.3 Einstein Telescope

Following the analysis developed in [43], the authors expect that the ET will observe $N = 10^3$ SS events over a three year observation period. The redshift probability distribution function for the SS events is given by

$$f(z) = \frac{4\pi \mathcal{N} r(z) d_L^2(z)}{H(z)(1+z)^3}, \tag{4.25}$$

where $\mathcal{N}$ is a normalization constant used that takes the value required to ensure that $f(z)$ is normalized to unity.

In the previous equation, $z_{\max}$ is the maximum redshift at which we expect that the ET is able to measure an event with a signal-to-noise ratio above 8, which we take to be $z_{\max} = 2$. The lower cutoff in the integral, $z_{\min}$, is the lowest redshift which requires a model of the local flow of the emitting source, which we take to be $z_{\min} = 0.07$.

The function $r(z)$ is called the coalescence rate at redshift $z$, and is given by

$$r(z) = \begin{cases} 1 + 2z & \text{if } 0 \leq z \leq 1 \\ (15 - 3z)/4 & \text{if } 1 < z < 5 \\ 0 & \text{if } z > 5 \end{cases} \tag{4.26}$$

Since we have set an upper and lower cutoff to the observations obtained by the ET, the previous function will only be considered inside this redshift range, between $z_{\min}$ and $z_{\max}$, instead of between $z = 0$ and $z = 5$.

As for the error, the estimated total error as a function of redshift for the ET is considered to be of the form

$$\sigma_{\text{ET}}^2 = \sigma_{\text{inst}}^2 + \sigma_{\text{lens}}^2, \tag{4.27}$$

where

$$\sigma_{\text{inst}} \approx \left(0.1449z - 0.0118z^2 + 0.0012z^3\right) d_L(z), \tag{4.28}$$

is the ET instrumental error and

$$\sigma_{\text{lens}} \approx 0.05z \, d_L(z), \tag{4.29}$$

is the error contribution due to lensing. Here we have chosen to neglect the error from the spectroscopic redshift measurements.

## 4.3 Bayesian Inference Methodology

In order to constrain the parameters for each model, we have performed a Bayesian analysis relying on Markov chain Monte Carlo (MCMC) methods. Specifically, we used PyStan [44], a Python interface to Stan [45], a statistical programming language which implements the No-U-turn sampler, a variant of the Hamiltonian Monte Carlo. The output was then analyzed using GetDist [46], to perform the corner plot, and ArviZ [47], which is a Python package that allows for an exploratory analysis of Bayesian models, namely by implementing posterior analysis, convergence diagnostics and model selection criteria.

For each combination of model plus dataset, we executed at least 4 independent chains, each of which with at least 2500 samples on the posterior distribution and at least 500 warm-up steps. To ensure that the chains converged, we use the $\hat{R}$ diagnostics, a convergence diagnostics introduced in [48], where it is claimed to have significant improvements when compared to other convergence diagnostics, namely the widely used Gelman-Rubin test. Roughly speaking, this diagnostics indicates how well the different chains are mixed. In this dissertation we have set an upper bound of $\hat{R} = 1.05$ in all chains, where $\hat{R} = 1$ is the ideal scenario.

The initial values are randomly sampled from a Gaussian distribution with mean around the expected value, which was the value used to generate the mock catalogs, and a standard deviation of roughly 10% of the corresponding mean for each parameter.

Finally, we consider what we believe to be weakly informative priors, which are given by a Gaussian distribution centered on the expected value and with a standard deviation of approximately two orders of magnitude larger than the previously mentioned value, to ensure a quasi flat prior around the region of interest.

## 4.4 Model Selection Criteria

Throughout this dissertation we will the require to differentiate between two models. To do this we need to evaluate the quality of the posterior predictions for a given model. This can be done by introducing a scoring rule that is referred to as the expected log pointwise predictive density (elpd) [49], which measures the quality of a model's fit when new data coming from the true data generating process is considered, after constraining the model with a given dataset. Mathematically the elpd is given by

$$\text{elpd} = \sum_{i=1}^{N} \int p_t(\tilde{y}_i) \ln p(\tilde{y}_i|y) d\tilde{y}_i \,, \tag{4.30}$$

where $N$ is the total number of events, $p_t(\tilde{y}_i)$ is the true distribution that generates the event $\tilde{y}_i$ and $p(\tilde{y}|y)$ is called the posterior predictive distribution which is defined as

$$p(\tilde{y}|y) = \int p(\tilde{y}_i|\theta) p(\theta|y) d\theta \,, \tag{4.31}$$

where $\theta$ is the vector of the model parameters, $p(\tilde{y}_i|\theta)$ is the probability of observing the event $\tilde{y}_i$ knowing the parameters $\theta$, i.e. the likelihood, and $p(\theta|y)$ is the posterior distribution.

In these very specific circumstances, we know the true distribution $p_t(\tilde{y}_i)$, given that these events are generated and do not correspond to real observations. However, if we are to assume to be oblivious to the data generating process, then we are required to approximate eq. (4.30).

The method that we will use here is referred to as leave-one-out cross-validation (LOO-CV), which consist of fitting our model to $N-1$ events of our dataset, and evaluate how likely is it that this model predicts the left out event. The elpd using LOO-CV is defined as [50]

$$\text{elpd}_{\text{LOO-CV}} = \sum_{i=1}^{N} \ln p(y_i|y_{-i}), \quad (4.32)$$

where $y_{-i}$ represents the dataset without the $i$-th observation and

$$p(y_i|y_{-i}) = \int p(y_i|\theta)p(\theta|y_{-i})d\theta, \quad (4.33)$$

gives the probability of predicting the event $y_i$ after constraining the model with the dataset $y_{-i}$.

However, LOO-CV methods are very computationally expensive, as one must repeat the full constraining procedure $N$ times, one for each omitted data point. To account for this problem, in [50] the authors present an efficient computation of LOO-CV, where a procedure is developed to estimate how the model would look like if constrained using $N-1$ events, from a single run of MCMC with the full sample of $N$ events. This approximation, which is referred to as Pareto smoothed importance sampling leave-one-out cross-validation (PSIS-LOO-CV), approximates the value of the elpd to

$$\text{elpd}_{\text{PSIS-LOO-CV}} = \sum_{i=1}^{N} \ln \left( \frac{\sum_{s=1}^{S} w_i^s p(y_i|\theta^s)}{\sum_{s=1}^{S} w_i^s} \right), \quad (4.34)$$

where $\theta^s$ is the $s$-th draw of the posterior distribution $p(\theta|y)$, $S$ is the total number of draws from the previous distribution and $\omega_i^s$ is the weight for the $s$-th draw and for the $i$-th event. The procedure used to compute the weights is detailed in [50]. In order to increase the readability, we will usually refer to the previous quantity simply as elpd and state that it was computed using PSIS-LOO-CV.

It is clear from the previous expression that a perfect model, i.e. one that would perfectly account for all observations, has an average of the elpd equal to zero, since $p(y_i|\theta^s) \approx 1$ for all events $i$, making the argument of the logarithm 1, and the sum over all events equal to 0. This means that, for the same dataset, one model is better than the other if its value of elpd is closer to zero. It is also relevant to point out that the higher the number of observations, the lower the expected value of the average of the elpd, since the sum will always include non-positive numbers.

The value for PSIS-LOO-CV estimate of the elpd, as well as its corresponding standard deviation, is implemented in ArviZ [47], a Python package for Bayesian inference, which was used throughout this dissertation.

## 4.5 Catalog Selection Criteria

To provide statistical confidence that the SSs catalogs used throughout this analysis are representative of a wide range of outcomes, we generate several catalogs, for each observatory, and study three representative cases: the best, the median and the worst.

The criteria used to categorize each catalog was based on an how small the multiplication of the $1\sigma$ region for all parameter is. This value, is formally defined as

$$\Delta_{\text{tot}}^2 \equiv \prod_{i=1}^{N} \sigma_{\theta_i}^2 \,, \tag{4.35}$$

where $\sigma_{\theta_i}$ is the standard deviation of the parameter $\theta_i$ and $N$ is the total number of parameters. For a given model the value of $\Delta_{\text{tot}}^2$ dictates how much constraining power a given dataset has, such that the smaller its value, the more certain we are about the values of the parameters.

We refer to the best catalog when we speak of the best possible (generated) outcome, whereas the worst catalog represents the worst possible (generated) outcome and the median catalog has the median constraining power of all of the generated mock catalogs.

# Chapter 5

# $f(Q)$ Cosmology with a $\Lambda$CDM Background

In this chapter, we will study the most general $f(Q)$ modified gravity cosmological model which features a $\Lambda$CDM background where the differences arise in the propagation of perturbations, namely by introducing an effective gravitational constant. This model introduces only one additional free parameter, $\alpha$, which when set to zero makes this model fall back to $\Lambda$CDM. Our main goal here is to assess the constraining power that both current and future GW observatories will have in setting the boundaries on the parameter $\alpha$, by resorting to SS mock catalogs.

This model has been the subject of study in previous original work carried out by the author in [13]. This model has also been addressed in [51], where it has been shown to alleviate the $\sigma_8$ tension present in $\Lambda$CDM and in [52], where an analysis using scalar angular power spectra, matter power and GWs propagation was developed. Similar models were also briefly studied in [26], a more general model of the form $f(Q) = Q + \alpha Q^n$ was studied in [53] and models of similar form were constrained using observational data in [54, 55].

## 5.1 The Model

In order to ensure that our cosmological model of $f(Q)$ has a $\Lambda$CDM background, we take the modified first Friedmann equation, presented in eq. (3.20), and set the right hand side to be equal to $H^2$. By making use of the relationship between the non-metricity scalar and the Hubble function, $Q = 6H^2$, we can then write

$$Qf_Q - \frac{1}{2}f - \frac{Q}{2} = 0, \tag{5.1}$$

with, as shown in [7], the most general solution given by

$$f = Q + \alpha\sqrt{Q}, \tag{5.2}$$

where $\alpha$ is an integration constant.

We will consider a universe which is composed of ordinary and dark matter, radiation and a cosmological constant. However, since the aim of this chapter is solely to forecast this model using SSs, we can safely neglect radiation, as these objects are not expected to be detected at very high redshifts. Since the background is the same as ΛCDM, we can make use of conservation laws and the Hubble function now reads

$$H = H_0\sqrt{\Omega_m(1+z)^3 + 1 - \Omega_m}\,. \qquad (5.3)$$

By inserting the specific form for this model, presented in eq. (5.2), in the equation for the luminosity distance of GWs in $f(Q)$, given in eq. (3.32), the measured luminosity distance of a GW is

$$d_{\text{GW}}(z) = \sqrt{\frac{2\sqrt{6}+\alpha}{2\sqrt{6}+\alpha/E(z)}}\, d_L(z)\,, \qquad (5.4)$$

where $\alpha$ has been re-defined to be in units of $1/H_0$ and $E(z) \equiv H(z)/H_0$.

Looking at the previous equation, one can see that there is a singularity at $\alpha = -2\sqrt{6}E(z)$. To ensure that the value for the luminosity distance for GWs is strictly physical (i.e. a real positive number for all redshifts), we must require that the value of $\alpha$ must have a lower bound at $\alpha = -2\sqrt{6}$.

## 5.2 Forecasts using Standard Sirens

By looking at the expression for the luminosity distance of gravitational waves, as can be seen in eq. (5.4), we can see that this model features a degeneracy between two of its parameters, $\alpha$ and $\Omega_m$. In order to avoid problems with our sampler, we make use of the Pantheon sample with the marginalized SnIa likelihood, developed in section 4.1, to fix the value of $\Omega_m$. As such, besides the SSs events, the Pantheon sample is present at all times.

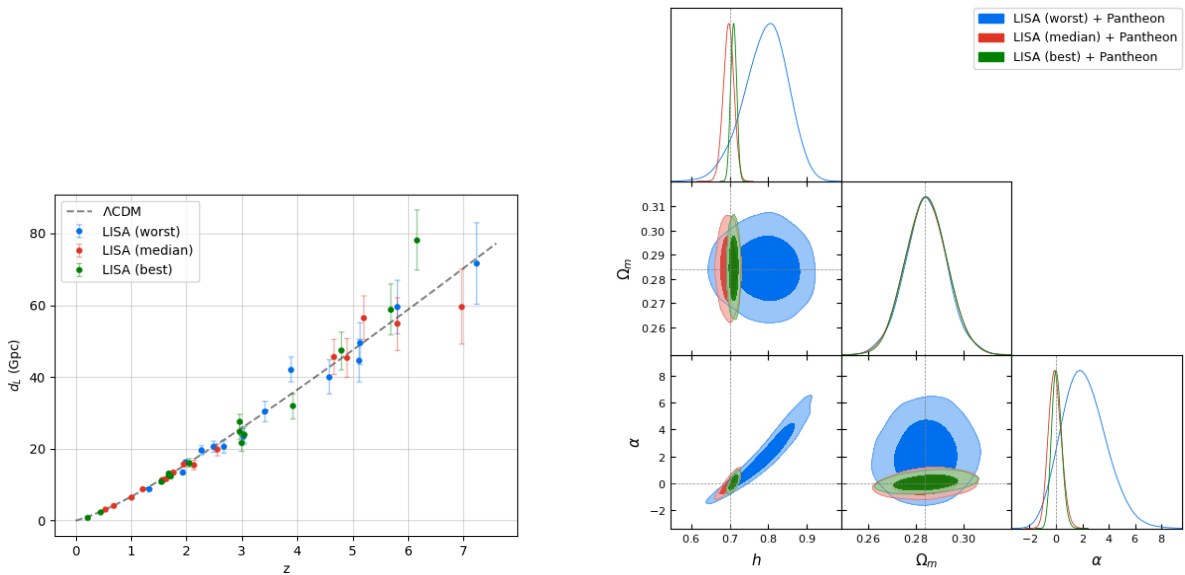

Figure 5.1: The best, median and worst LISA catalogs represented in the luminosity distance versus redshift plane (left) and the corresponding constrains with the Pantheon sample set on the model given by eq. (5.2) (right). The fiducial ΛCDM cosmological model is plotted as a dashed gray line.

Starting the forecast with LISA, we present in the left panel of fig. 5.1 the best, median and worst LISA catalogs, which are rated as previously described in chapter 4, on the luminosity distance versus redshift plane. On the right panel of the same figure, we show the corner plot illustrating the constrains on the parameters obtained by a joint analysis of the same catalogs, together with the Pantheon SnIa data.

By looking at the constraints in the right plot of fig. 5.1, we can see that the worst catalog features significantly worst results when compared to both the median and best catalogs, which in turn are very similar when compared to each other. If we then look at the redshift distribution for the events presented in the left plot of fig. 5.1, we can see a pattern: the best catalog is the one which features events with the lowest redshifts, followed by the median catalog, and finally the worst catalog. From this fact, we infer that LISA increases the quality of its constraints when the corresponding catalog features plenty of low redshift events.

Considering that, out of the 15 generated catalogs, only 2 provided error bars similar to the ones obtained in the worst catalog, we expect that the odds of obtaining a bad LISA catalog to be low.

Focusing our attention on the ET, we noticed that all of the 5 generated catalogs provide similar constraints. This is expected, as the number of events which we are considering for this observatory, 1000 to be precise, is large enough to match the underlying statistical distribution. As such, we decided only to consider one ET catalog, which we take as being representative of a whole range of outcomes for this observatory.

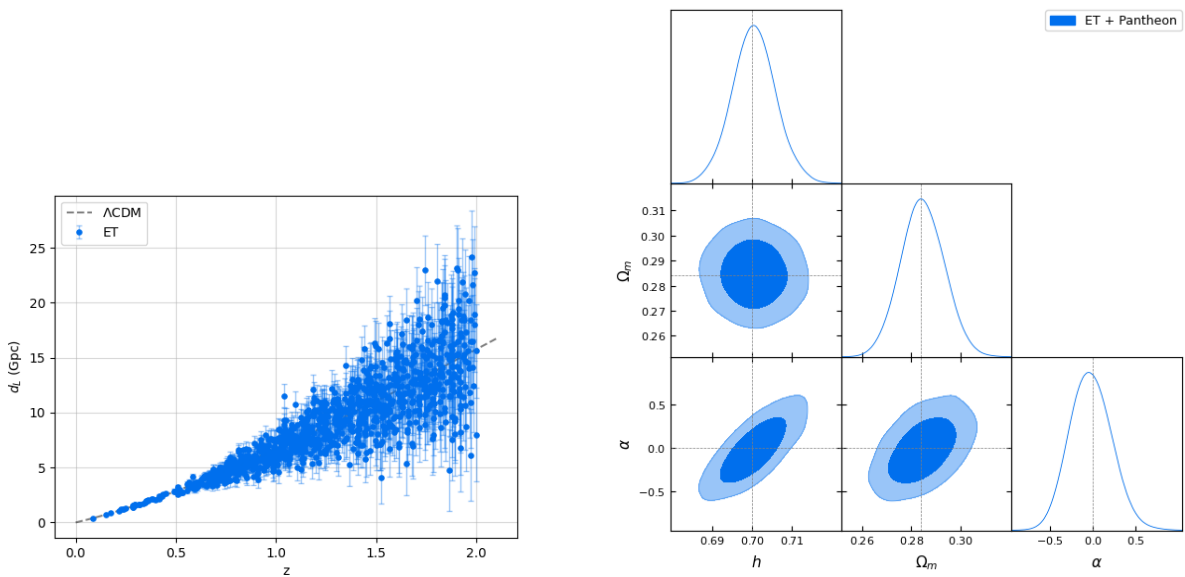

Figure 5.2: The considered ET catalog represented in the luminosity distance versus redshift plane (left) and the corresponding constrains with the Pantheon sample set on the model given by eq. (5.2) (right). The fiducial ΛCDM cosmological model is plotted as a dashed gray line.

The single ET catalog is presented in the left plot of fig. 5.2, in the luminosity distance versus redshift plane, and the corresponding constraints for that catalog, when mixed with SnIa obtained from the Pantheon sample, are presented in the right plot of fig. 5.2.

If we are now to compare the constraints set by the ET with the ones set by the best catalog LISA, where both catalogs are presented in the left plot of fig. 5.3 and the corresponding constraints are presented in the right plot of fig. 5.3.

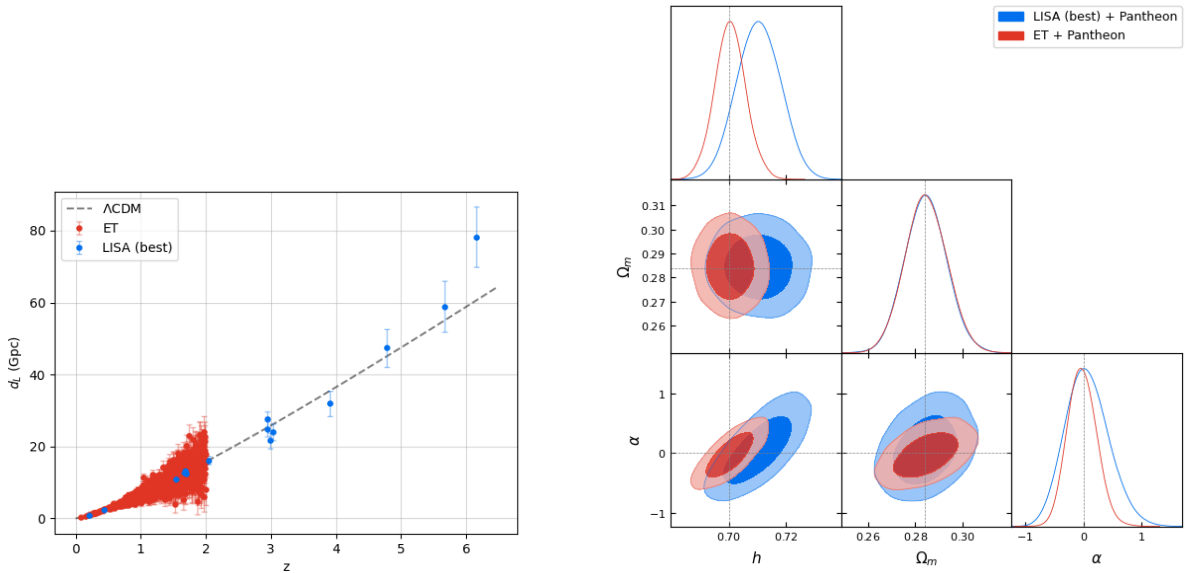

Figure 5.3: The considered ET catalog and the best LISA catalog represented in the luminosity distance versus redshift plane (left) and the corresponding constrains with the Pantheon sample set on the model given by eq. (5.2) (right). The fiducial ΛCDM cosmological model is plotted as a dashed gray line.

By computing the area of the $1\sigma$ region we can see that the best LISA catalog has a region which is approximately 4 times bigger when compared to the one given by the ET. The fact that the ET will operate at significantly lower redshift when compared to LISA, yet is able to provide better constraints, goes hand in hand with our previous statement regarding LISA which is the fact that low redshift events are ideal to constrain this model. Granted, the ET will provide a significantly larger number of events when compared to LISA.

As for the LIGO-Virgo collaboration, the forecasts indicate that these observatories are not able to constrain this model. As such, we have decided to categorize each LIGO-Virgo catalog based on how well it complements the worst LISA catalog. The worst LISA catalog is chosen because both the best and median LISA catalogs do not show significant improvements when added with any of the LIGO-Virgo catalogs.

To understand why LIGO-Virgo itself is not able to set proper constraints on our model, we refer to fig. 5.4, where we can see a zoom of the events present in the low redshift regime, $z \in [0, 0.2]$, showcasing the best LIGO-Virgo catalog, as well as the best LISA catalog and the single ET catalog. This plot shows that, although there is a much larger number of events, the LIGO-Virgo error bars are significantly larger when compared to those of the ET and LISA.

The forecasts set by LIGO-Virgo, when joined together with the worst LISA catalog and the SnIa from the Pantheon sample, are presented in fig. 5.5. These results reveal that if the events observed by LISA resembles a worst LISA catalog, LIGO-Virgo is expected to significantly increase the quality of the constraints, and approximate its constraining power to that of a median LISA catalog.

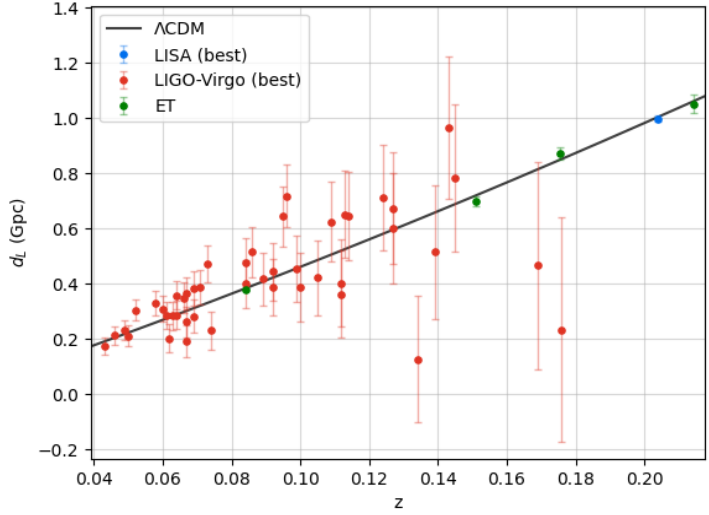

Figure 5.4: Luminosity distance, in Gpc, as a function of redshift, for the best LISA catalog, the best LIGO-Virgo catalog and the single ET catalog. The ΛCDM luminosity distance is plotted as a solid gray line.

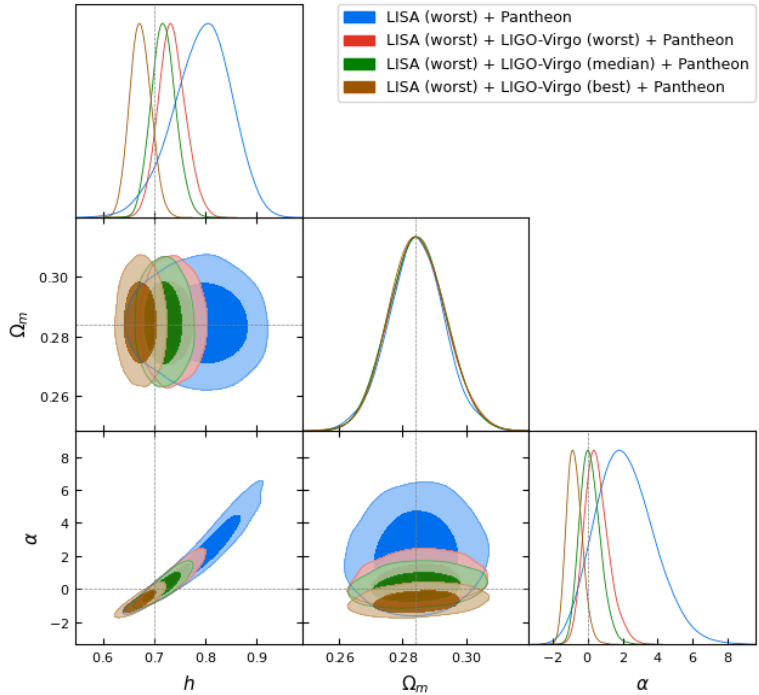

Figure 5.5: Constraints set by the worst, median and best LIGO catalogs, when mixed together with the worst LISA catalog and data from the Pantheon dataset, for the model given by eq. (5.2). The worst LISA catalog with Pantheon is also shown for comparison. Dotted lines represent the fiducial ΛCDM values.

In an attempt to further increase the quality of our constraints with the available catalogs, we added each of the LISA catalogs with the single ET catalog, where only the best LISA catalog showed non-negligible improvements. As for the data of LIGO-Virgo when mixed with the ET, no significant improvements were observed. These results were to be expected given by both the high quality of the constraints and the number of events measured by the ET.

The results set by LISA show that, even though the luminosity distance for this model deviates from that of $\Lambda$CDM as the redshift increases, the error bars increase faster, in such a way that high redshift events are less useful to constrain this model. On the other hand, were we to observe an event at the very low redshifts with a larger error bar, as was clearly seen for the case of LIGO-Virgo, then we would be unable to set proper constraints. We also saw that this is also true for LISA, since in this regime the current model is practically indistinguishable from $\Lambda$CDM and, to further complicate matters, the observations are more sensitive to the peculiar velocities, consequently increasing the error bars of our measurements.

We noted that the value for the $1\sigma$ region for the parameter $h$, which we label as $\sigma_h$, is approximately linear with the $1\sigma$ region for the parameter $\alpha$, labeled as $\sigma_\alpha$. We would also like to note that the value for $\Omega_m$, as well as its error $\sigma_{\Omega_m}$, are the same for all catalogs regardless of the observatory considered. This is due to the usage of SnIa that fixes the value of $\Omega_m$. As such, without losing any information with respect to the quality of the constrains on the other parameters, we can categorize each catalog based on the value of $\sigma_\alpha$.

In table 5.1 we show the value of $\sigma_\alpha$ for the more relevant cases, by decreasing order of constraining power, as well as the relative size of each region with respect to the best expected outcome.

| Catalog | $\sigma_\alpha$ | Relative Size |
|---|---|---|
| ET | 0.25 | 1 |
| LISA (best) | 0.37 | 1.5 |
| LISA (worst) + LIGO-Virgo (best) | 0.44 | 1.8 |
| LISA (median) | 0.49 | 2 |
| LISA (worst) | 1.70 | 6.8 |

Table 5.1: Summary of the size of the $1\sigma$ region for the parameter $\alpha$, labeled as $\sigma_\alpha$, for the more relevant cases, ordered by decreasing constraining power. The relative size of $\sigma_\alpha$ for each case compare to the best expected outcome is also presented.

Quantitatively, table 5.1 shows that we can expect that the ET will be able to measure the value of $\alpha$ with a $1\sigma$ region of 0.25. Then, our most optimistic result for LISA shows that it will be able to set the bounds on $\alpha$ on a region which is 0.5 larger than that of the ET. If instead we are to observe a bad LISA catalog, then we expect to have a $1\sigma$ region which is almost 7 times worst when compared to the ET. However, if we are to measure the equivalent of the best case for LIGO-Virgo, the quality of those constrain will be able to improve enormously to reveal only 0.8 worst when compared to the ET. This last case, although features a bad LISA catalog, it is slightly better when compared to the median LISA catalog, which will be 2 times worst compared to the ET.

## 5.3 Summary

In this chapter, we have considered the most general model of $f(Q)$ gravity that mimics a ΛCDM background, which introduces one additional free parameter. Due to this choice of background, departures from ΛCDM only arise at the perturbative level.

We forecast this model using SS mock catalogs generated for LISA, ET and LIGO-Virgo. Additionally, due to a degeneracy between two of the model parameters, each mock catalog was constrained in conjunction with SnIa from the Pantheon sample.

The LISA catalog which provided the best constraints is the one which features more events at low redshifts, while the worst catalog is composed mostly of high redshift events, and the median is somewhere in between. This result indicates that, even though our model deviates more from ΛCDM as the redshift increase, the error bars increase faster, making high redshift observations less useful to provide constraints. Based on all of the generated LISA catalogs, a catalog which provides similar constraints to the worst LISA catalog is not very likely to be observed.

For the ET it was observed that all catalogs provide similar constraints and as such only had to consider one catalog to be representative of the whole set. This is expected since each catalog features 1000 SS events, a number which is large enough to represent the underlying probability distribution function. On top of this consistency, the ET performs slightly better than the best LISA catalog.

As for LIGO-Virgo, we observed that a dataset consisting of 50 SSs events are not enough to provide meaningful constraints. Instead, we categorized each LIGO-Virgo catalog based on how well it would complement the worst LISA catalog. We showed that one can rely on future events obtained by LIGO-Virgo to improve the quality of the constraints, meaning that, in the future, if we obtain a catalog similar to that of a worst LISA catalog, we can use data from LIGO-Virgo to approximate it to have the quality of a median LISA catalog. By contrast, for both the best or the median LISA catalogs, none of the LIGO-Virgo catalogs made significant improvements on the results.

We can provide small improvements to the constraints set by the ET using the best LISA catalog. By contrast, the median and the worst LISA catalog, as well as any of the LIGO-Virgo catalogs, make no significant improvements when constrained together with the ET.

We observed that the size of the $1\sigma$ region for $\alpha$ has a quasi-linear relationship with the size of the $1\sigma$ region for $h$. Due to the fact that the parameter $\Omega_m$ was fixed by the usage of SnIa, we were able to rate the quality of the constrains of each catalog based only on the size of $\sigma_\alpha$, without losing information regarding the quality of the constrains of the other parameters.

In table 5.1 we summarize the forecasts of the constrains set on the parameter $\alpha$ for the more relevant cases we have studied. Quantitatively, we show that the ET will be able to set a $1\sigma$ region for the parameter $\alpha$ of size 0.25, whereas the best possible scenario for LISA shows a region which is about 0.5 larger. The median LISA catalog showed a $1\sigma$ region which is 2 times larger, again compared with the constrains set by the ET, while the worst possible case expected for LISA revealed a region which is almost 7 times larger. Fortunately, the data from LIGO-Virgo is expected to improve the quality of the constrains, which in a best case scenario is expected to improve the size of the $1\sigma$ region from 7 times larger to only 0.8 times worst when compared to the ET, beating the result set by a median LISA catalog.

# Chapter 6

# $f(Q)$ as Dark Energy

One of the major interests behind the study of modified gravity is to provide a theoretical explanation to dark energy. Given that the model studied in the previous chapter is not able to explain the accelerated expansion of the universe without relying on a cosmological constant, in this chapter we will study an $f(Q)$ cosmological model where the modification to gravity accounts for its late time evolution. This model introduces a multiplicative term in the action that only modifies gravity in the low redshift regime. This extension of gravity does not add any extra degrees of freedom, given that it plays the role of a cosmological constant which, using the conservation equation, can be expressed as a function of both matter and radiation.

First, we will perform a dynamical system analysis[1] in order to narrow down the region in parameter space which provides viable cosmological models. Afterwards, we will use the SS mock catalogs in order to distinguish this model from ΛCDM. The usage of SnIa will also be employed in order to assist in the model selection process, given that SS alone are inconclusive.

This model was originally proposed in [57], where it was shown to be statistically equivalent to ΛCDM, for tests using SnIa, redshift space distortions and cosmic chronometers. Additionally, it has been studied in [58], where it was shown to pass Big Bang nucleosynthesis constrains, showing that this modification to gravity is able to capture the accelerated expansion of the universe at late times, without modifying the early ages of the universe in a way that is compatible with observational data.

## 6.1 Dynamical System Analysis

As initially proposed in [57], we consider a specific form of the function $f(Q)$ that replicates late time acceleration. We do this by introducing an extra term in the action that only impacts the low redshift regime, which caries an additional parameter $\lambda$, that plays the role of the cosmological constant. As for the energy matter content in the universe, we are dropping the assumption of the presence of a cosmological constant, and we consider a universe which is permeated solely by a perfect fluid composed of matter and radiation.

The action for this theory reads

$$f = Qe^{\lambda Q_0/Q}, \tag{6.1}$$

---
[1] The main reference used for the theory behind the dynamical systems analysis was [56].

where $Q_0$ is the non-metricity scalar, $Q = 6H^2$, evaluated at the present day.

Rewriting the modified first Friedmann equation for a generic cosmological model of $f(Q)$ gravity, presented in eq. (3.20), by making use of the relative abundances, which were defined in eq. (3.18), we obtain

$$1 = \frac{1}{2}\frac{f}{Qf_Q} + \frac{1}{2f_Q E^2}\left(\frac{\Omega_m}{a^3} + \frac{\Omega_r}{a^4}\right). \tag{6.2}$$

Computing the value of $f_Q$, and making use of the relationship between the non-metricity scalar and the Hubble function in eq. (3.19), we obtain

$$f_Q = \left(1 - \frac{\lambda}{E^2}\right)e^{\lambda/E^2}, \tag{6.3}$$

which can be inserted back into eq. (6.2) to give

$$1 = \frac{e^{-\lambda/E^2}}{E^2}\left(\frac{\Omega_m}{a^3} + \frac{\Omega_r}{a^4}\right) + \frac{2\lambda}{E^2}. \tag{6.4}$$

If we define the quantities

$$x_1 \equiv \frac{\Omega_m}{E^2 e^{\lambda/E^2} a^3}, \qquad x_2 \equiv \frac{\Omega_r}{E^2 e^{\lambda/E^2} a^4}, \qquad x_3 \equiv \frac{2\lambda}{E^2}, \tag{6.5}$$

we can find a closed system of the form

$$x_1 + x_2 + x_3 = 1. \tag{6.6}$$

From the previous equation, we can see that the coordinate $x_1$ is related to the evolution of the matter density in the universe, whereas $x_2$ is related to the radiation density and $x_3$ to the contribution of our modification to gravity. Additionally, this equation shows that one of our coordinates in phase space is fully determined by the value of the other two. As such, without loss of generality, we choose to work only with $x_1$ and $x_2$.

In order to understand the evolution of the background dynamics of this universe, we introduce the number of e-folds, which is mathematically expressed as

$$N \equiv \ln a, \tag{6.7}$$

which we will now use as a proxy for time in our dynamical system. Computing the derivatives of the coordinates in phase space with respect to the number of e-folds, which from now one we will denote with a prime, reveals that

$$x_1' = -x_1\left(-(1 + x_1 + x_2)\frac{E'}{E} + 3\right), \qquad x_2' = -x_2\left(-(1 + x_1 + x_2)\frac{E'}{E} + 4\right), \tag{6.8}$$

where the factor $E'/E$ can be expressed as a function of both $x_1$ and $x_2$ by

$$\frac{E'}{E} \equiv \frac{3x_1 + 4x_2}{2 - (x_1 + x_2) + (x_1 + x_2)^2}. \tag{6.9}$$

By comparing $x_1'$ and $x_2'$ we see a symmetry of our system: both coordinates change in a very similar fashion with respect to the number of e-folds, where the only difference is a factor of 3 in $dx_1/dN$ transforming into a 4 in $dx_2/dN$. This difference is due to the evolution of each component, given that $x_1$ is related to the matter density, which scales with $a^{-3}$, whereas $x_2$ is related to the evolution of radiation, and therefore scales with $a^{-4}$.

### 6.1.1 Fixed Points

In order to characterize the dynamics of our system in phase space, it is useful to compute the fixed points and their corresponding stability. A fixed point of a dynamical system is characterized as any given point in phase space where the system does not change with time. Whether a given system will naturally tend towards or away from this fixed point is given by its stability. These points are of great interest because they allow us to qualitatively characterize the trajectories of the system in phase space, in a small neighborhood around them. Computing the fixed points is done by setting the equations which rule the dynamical system to zero.

In order to compute the stability of a given fixed point, we will make use of linear stability theory. This is done by computing the eigenvalues of the Jacobian matrix when computed in each of the fixed points. The stability of each fixed point is characterized by the sign of its corresponding eigenvalues as follows:

- Stable: all of the eigenvalues have negative real part;

- Saddle: at least two eigenvalues have real part with opposite signs;

- Unstable: all of the eigenvalues have positive real part.

Computing the fixed points for the dynamical system, as well as the corresponding stability for each, reveals 3 distinct points, which are presented in table 6.1.

| Fixed Point | Type | $x_1$ | $x_2$ | $x_3$ | Stability | Eigenvalues |
|---|---|---|---|---|---|---|
| I | $\lambda$ dominated | 0 | 0 | 1 | Stable | (-4, -3) |
| II | Matter dominated | 1 | 0 | 0 | Saddle | (3, -1) |
| III | Radiation dominated | 0 | 1 | 0 | Unstable | (4, 1) |

Table 6.1: Fixed points and corresponding stability for the dynamical system.

It is important to note that the previous procedure is not always valid, and only works if all of the eigenvalues obtained have a non-zero real part, otherwise different and move advanced techniques must be used in order to asses the stability of a fixed point. Fortunately all of the eigenvalues for each fixed point in this system have non-zero real part, meaning that we can make use of linear stability theory to characterize the orbits around each of the fixed points.

In the following subsections, we will study the behavior of the dynamical system on each of its fixed points.

**Fixed Point *I***

The first fixed point we will study, which we have labeled as *I*, is the only stable fixed point in our system. Located at $x_1 = 0$ and $x_2 = 0$, by making use of eq. (6.6), it immediately follows that $x_3 = 1$. This in turn means that for this fixed point we have

$$E^2 = 2\lambda, \tag{6.10}$$

which means that in this fixed point the value of $\lambda$ will always be a positive number.

If we are to integrate the previous equation, knowing that the Hubble constant is positive, we obtain the evolution of the scale factor as a function of time

$$a = e^{\sqrt{2H_0^2 \lambda} t}, \tag{6.11}$$

which means that we have an exponentially accelerated expansion. As such, this fixed point corresponds to a point of eternal inflation.

**Fixed Point *II***

The fixed point *II*, which from the stability point of view is a saddle point, is stable along the direction which connects this fixed point to the fixed point *III*, and unstable in the direction perpendicular to that line. Given that it is located at $x_1 = 1$ and $x_2 = 0$, it implies by eq. (6.6) that $x_3 = 0$, which in turn implies that $\lambda = 0$.

If we remember our model of $f(Q)$, presented in eq. (6.1), it is easy to see that if we set $\lambda = 0$ we fall back to the STEGR, but not to $\Lambda$CDM, since our model does not include dark energy, meaning that we fall back to a cold dark matter (CDM) model.

By making use of the condition $x_2 = 0$ it follows that $\Omega_r = 0$, implying that this fixed point corresponds to a matter dominated CDM model.

**Fixed Point *III***

The final fixed point in our system, fixed point *III*, is unstable and located at $x_1 = 0$ and $x_2 = 1$. Just like fixed point *II*, it is also located in a region where $\lambda = 0$, meaning that this $f(Q)$ model falls back to a CDM model.

By making use of the condition $x_1 = 0$ it follows that $\Omega_m = 0$, which means that this fixed point corresponds to a radiation dominated CDM universe.

### 6.1.2 Trajectories

Making use of numerical tools to solve the equations of motion for this dynamical system, we obtained the trajectories in phase space, which we show in fig. 6.1, along with the fixed points. The velocity vectors are represented by the black arrows, the stable fixed point, which we refer in text as *I* is marked in green, the saddle fixed point which we refer to as fixed point *II* is marked in orange and the unstable fixed point, referred to as *III*, is marked in red.

Resorting to fig. 6.1, we can visually identify three disjoint regions[2] in phase space, which are categorized based on the sign of the parameter $\lambda$ as follows:

- Region *A*: region with $\lambda > 0$, colored in light blue, which corresponds to the triangle $x_1 + x_2 < 1$. This is where fixed point *I* (the stable fixed point) is located, and therefore attracts all trajectories which are inside that triangle.

- Region *B*: line with $\lambda = 0$, in light brown, corresponding to the case $x_1 + x_2 = 1$, which is a CDM cosmological model. This is where both the fixed point *II* (saddle) and *III* (unstable) are located, with the trajectories flowing from the unstable point towards the saddle point.

- Region *C*: region $\lambda < 0$, marked in dark blue, which corresponds to the plane $x_1 + x_2 > 1$. It includes no fixed points, and all trajectories diverge towards infinity.

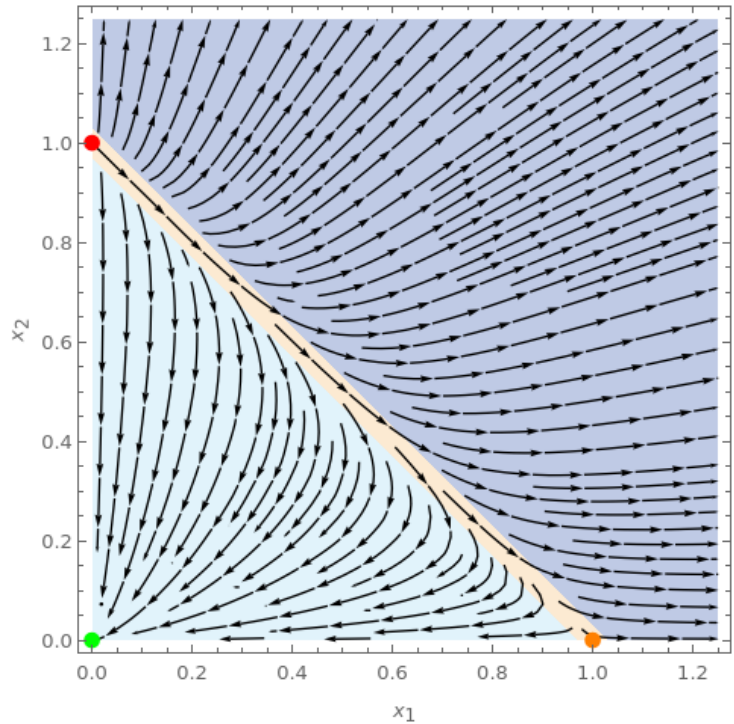

Figure 6.1: Stream plot of the phase space for the dynamical system being studied where the fixed point *I* (stable) is represented in green, fixed point *II* (saddle) in orange and fixed point *III* (unstable) is in red. The light blue marks the region where $\lambda > 0$, the light brown region corresponds to $\lambda = 0$ and the dark blue region to $\lambda < 0$.

We can see that for any initial state of the universe which falls inside region *A*, it follows that the final state of the universe will always be fixed point *I*, which corresponds to a eternal inflation.

However, if the initial state of the universe happens to be in the line corresponding to region *B*, the universe will repel away from the fixed point *III*, a radiation dominated epoch, and flow towards fixed point *II*, a matter dominated epoch, which will be the ultimate fate of the universe.

---

[2]The fact that there are trajectories which seem to flow away from the region in light brown, which corresponds to the case $\lambda = 0$, towards the light and dark blue region, which correspond to $\lambda > 0$ and $\lambda < 0$ respectively, is merely due to the fact that the initial conditions are not exactly on top of the line corresponding to $\lambda = 0$.

As for region *C*, as it was mentioned before, there are no fixed points and all trajectories diverge towards infinity. This means that both $x_1$ and $x_2$ will increase perpetually implying that the scale factor must decrease and asymptotically approach zero, which physically corresponds to a universe with a big crunch.

Current observations of the state of the universe suggest that we are headed towards a scenario of never-ending expansion. From these the three disjoint regions only the first is able to agree with observations. As such, this forces us to assume that $\lambda$ must be positive, which is in agreement to what was obtained in [57], where this model was constrained using current observations from different sources of data.

If we insert the function $f(Q)$ we are considering for this model, introduced in eq. (6.1), in the modified first Friedmann equation, presented in eq. (3.20), we can relate the value of $\lambda$ with both $\Omega_m$ and $\Omega_r$ as

$$(E^2 - 2\lambda)e^{\lambda/E^2} = \Omega_m(1+z)^3 + \Omega_r(1+z)^4, \tag{6.12}$$

which when considered for the present day and performing some rearrangements yields

$$\left(\lambda - \frac{1}{2}\right)e^{\lambda - 1/2} = -\frac{\Omega_m + \Omega_r}{2e^{1/2}}. \tag{6.13}$$

The previous system only possesses a solution if we require that the sum of the densities to be lower than $2e^{-1/2}$. This is not immediately ensured because, unlike what happens in $\Lambda$CDM where the sum of the densities is always equal to 1, here the first Friedmann equation is modified, setting no boundaries on the sum of the densities today. Taking this upper bound into consideration, and also the fact that each of the densities are always positive, then the right hand side of the previous equation obeys the inequality

$$-\frac{1}{e} \leq -\frac{\Omega_m + \Omega_r}{2e^{1/2}} < 0, \tag{6.14}$$

and thus we are able to express the possible solutions for $\lambda$ as

$$\lambda_0 = \frac{1}{2} + W_0\left(-\frac{\Omega_m + \Omega_r}{2e^{1/2}}\right), \qquad \lambda_{-1} = \frac{1}{2} + W_{-1}\left(-\frac{\Omega_m + \Omega_r}{2e^{1/2}}\right), \tag{6.15}$$

where $W_0$ and $W_{-1}$ are the main and the $-1$ branch of the Lambert function respectively.

We can see that there are two possible solutions for the value of $\lambda$. However, as we have previously discussed, the value of $\lambda$ must be positive in order for the model to agree with observations. Looking at both branches, we can see that the value of $\lambda_{-1}$ is strictly negative, leaving $\lambda_0$ as the only physically meaningful solution to this problem. Additionally, we see that the value of $\lambda_0$ takes negative values when $\Omega_m + \Omega_r > 1$. This means that, in order to agree with observations at all times, the sum of the densities must satisfy the constraint $\Omega_m + \Omega_r < 1$, which is familiar to us given that this is what one expects from $\Lambda$CDM.

## 6.2 Model Selection using Standard Sirens

In this section we will be using SS mock catalogs to see whether future data will be able to distinguish this model from ΛCDM apart. It is important to note that, just like in the previous model, we neglect the effect of radiation from now on, on the basis that SS are not expected to take place in epochs where radiation plays an important role.

Inserting the function $f(Q)$ specific for this model, given in eq. (6.1), in the generic form for the luminosity distance of GWs for a cosmological model based on $f(Q)$ gravity, presented in eq. (3.32), we obtain

$$d_L^{(\text{GW})}(z) = \sqrt{\frac{1-\lambda}{1-\lambda/E^2}} e^{\frac{\lambda}{2}(1-1/E^2)} d_L(z) \,. \tag{6.16}$$

Using the model selection criteria introduced in section 4.4, we take the best LISA catalog, as well as the single ET catalog, and compute the corresponding values of the elpd using PSIS-LOO-CV for this model and ΛCDM. The results for the model selection criteria are presented in table 6.2, whereas the best-fit values are shown in table 6.3. Additionally, the best-fit for the model presented in eq. (6.1) and ΛCDM when using data from the ET are presented in fig. 6.2.

| Model | elpd | |
|---|---|---|
| | LISA (best) | ET |
| $f(Q)$ as Dark Energy | $-19.58 \pm 7.17$ | $-1720.99 \pm 36.50$ |
| ΛCDM | $-19.76 \pm 7.19$ | $-1721.00 \pm 36.52$ |

Table 6.2: The average of the elpd, computed using PSIS-LOO-CV, for the model given by eq. (6.1) and ΛCDM, for the best LISA catalog and the single ET catalog.

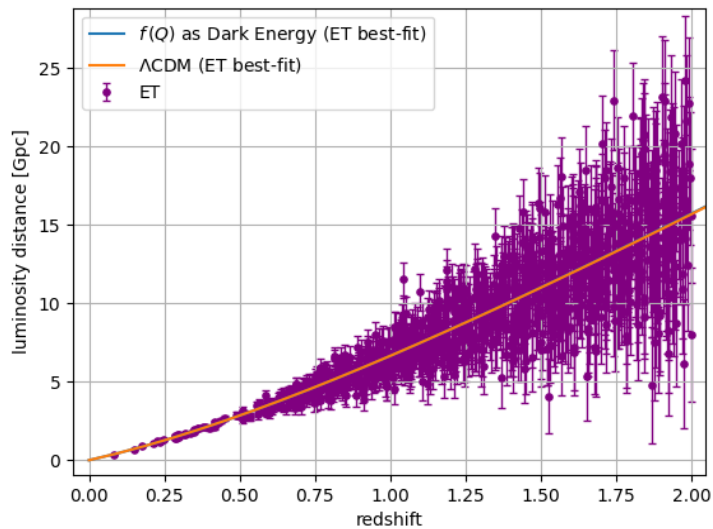

Figure 6.2: The best-fit luminosity distance curve, in Gpc, versus the redshift for model eq. (6.1) and ΛCDM, when using data from the ET.

| Model | $h$ | | $\Omega_m$ | |
|---|---|---|---|---|
| | LISA (best) | ET | LISA (best) | ET |
| $f(Q)$ as Dark Energy | $0.7118 \pm 0.0092$ | $0.6989 \pm 0.0057$ | $0.205^{+0.021}_{-0.025}$ | $0.257 \pm 0.016$ |
| $\Lambda$CDM | $0.7127 \pm 0.0093$ | $0.6995 \pm 0.0058$ | $0.236^{+0.023}_{-0.026}$ | $0.290 \pm 0.017$ |

Table 6.3: The best fit values and corresponding $1\sigma$ region for both $h$ and $\Omega_m$, for the model given by eq. (6.1) and $\Lambda$CDM, for the best LISA catalog and the single ET catalog.

Based on the results presented in table 6.2 we can conclude that the two models behave similarly when it comes to predicting held-out data. Looking at fig. 6.2 we can see that there is an overlap between the best-fits of the luminosity distance of both models, and that both replicate the original distribution used to create the SS events, meaning that both fit the data well. This goes to show that this model is essentially indistinguishable from $\Lambda$CDM.

The lower values of the elpd for the ET when compared to LISA are explained due to the higher number of events the former has when compared to the later, as discussed in section 4.4.

From table 6.3 it is possible to see that, for a given catalog, the best fit values of $\Omega_m$ for both $\Lambda$CDM and for the model we are considering, agree on the value of $h$, but do not agree on the value of $\Omega_m$. It is important to note that the value of $\Omega_m$ obtained for LISA and for the ET are lower than expected. However, this was merely a statistical fluctuations, as when consider the other LISA catalogs there are others which feature a larger value of $\Omega_m$.

All other mock catalogs which we have applied model selection criteria to revealed, with different degrees of precision, results that agree with the conclusions that we have just discussed. This is something we expected, since if we are unable to distinguish between two models with the best catalogs, then it is not expected that any of the other catalogs are able to differentiate between the two.

We would also like to note that the constraining power of a given catalog for the previous model is the almost the same as when considering this catalog, meaning that a good (or bad) catalog for the previous model is also a good (or bad) catalog for this model.

### 6.2.1 Varying the Number of Events

Given that we are unable to distinguish between the two models with the mock catalogs we have generated, we then asked ourselves whether this was a matter of precision, and decided to investigate whether it is possible to increase the number of SS events such that differences between this model and $\Lambda$CDM could be observed.

In order to minimize statistical fluctuations, for each number of events per catalog, represented by $N$, we generate $J$ different catalogs. Each catalog, which we index by the letter $j$ ranging from $j = 1$ to $j = J$, will then be used to compute the value of elpd using PSIS-LOO-CV. The final result is provided by averaging out the values obtained for all catalogs with the same $N$, propagating the uncertainties according to the standard rules of propagation of uncertainty.

For LISA we take $N = 5, 10, 15, 20, 25, 30$ events per catalog, and generate $J = 15$ different catalogs. For the ET we take $N = 500, 750, 1000, 1250, 1500, 1750$, and generate $J = 5$ catalogs. As for LIGO-Virgo, no realistic number of events were able to set constrains on this model.

The absolute value of the average of the elpd, computed by PSIS-LOO-CV as well as the corresponding error, as a function of the number of events per catalog $N$ obtained for LISA is represented in the left plot of fig. 6.3, whereas the results obtained for the ET are present on the right plot.

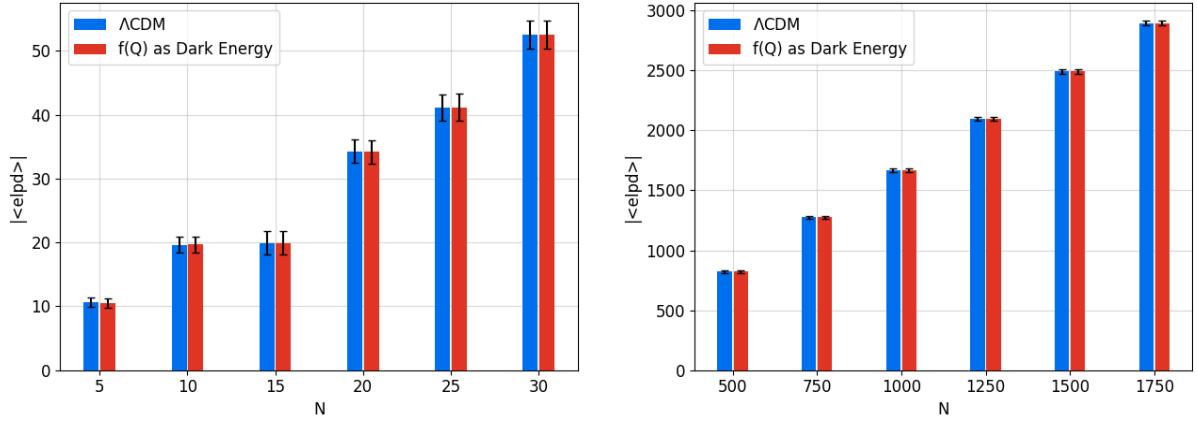

Figure 6.3: The absolute value of the elpd, computed by PSIS-LOO-CV, for both the model presented in eq. (6.1) and $\Lambda$CDM, when considering all catalogs with the same number of events per catalog. The plot on the left was made using data coming from LISA, whereas the one on the right uses data coming from the ET.

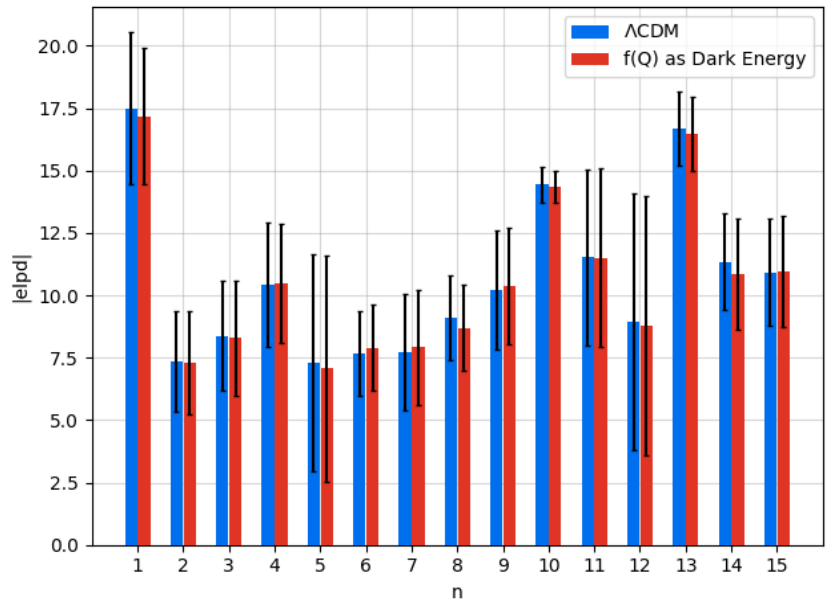

Figure 6.4: The absolute value of the elpd, computed by PSIS-LOO-CV, for both the model presented in eq. (6.1) and $\Lambda$CDM, for all LISA catalogs with $N = 15$ events per catalog.

The most important observation we can make is that, regardless of the value of $N$, and whether we are considering LISA or the ET, the value of the elpd tells us that these two models behave equally well on predicting held-out data. In fact, this equivalence is stronger, since this is something which happens on a per catalog basis. As an example we show in fig. 6.4 the value of the elpd for all LISA catalogs with $N = 15$.

We can therefore extrapolate that, regardless of the number of SS events that we are to consider, the luminosity distance curve for this model is able to closely match the luminosity distance curve for ΛCDM.

It is important to note that all values of the elpd are negative, as we have remarked in section 4.4. The absolute value is shown instead to increase the readability of the figures. The decreasing value of the elpd with increasing $N$ is explained using the same reasoning presented in section 4.4, which is that more events will always decrease the value of the elpd.

### 6.2.2 Using Type Ia Supernovae

The exercise in the previous section clearly shows that it is not possible to distinguish this model from ΛCDM based on SS alone, regardless of how many we expect to observe. We will now add current SnIa data and again apply model selection criteria to see whether this dataset prefers the model given by eq. (6.1) or ΛCDM. Because of the marginalization's performed on the likelihood for SnIa, in order to remove the degeneracy between $M$ and $H_0$, we are only able to constrain the value of $\Omega_m$ using this dataset.

Applying the model selection criteria, we present the values of the elpd computed using PSIS-LOO-CV in table 6.4. In fig. 6.5 we present the best-fit curves of the logarithm of the $H_0$ independent luminosity distance function as a function of redshift, $5\log D_L(z)$, as well as the events from the Pantheon sample.

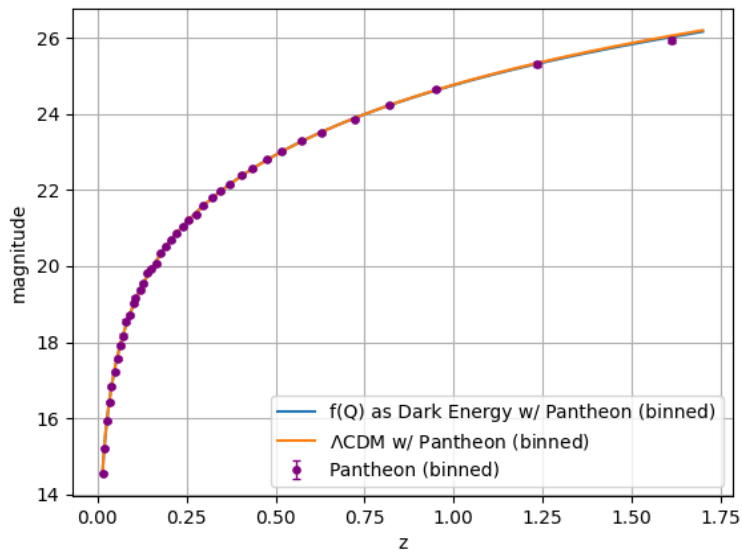

Figure 6.5: The best fit of the logarithm of the $H_0$ independent luminosity distance, $5\log D_L(z)$, versus redshift for the model in eq. (6.1) and ΛCDM, as well as the magnitude versus redshift of the Pantheon events.

From fig. 6.5 we can see that the best-fits for each model are essentially indistinguishable, and seem to be in agreement with the observations considered. As measured quantitatively by table 6.4, both models behave equally well on predicting held-out data, which means that SnIa are unable to distinguish this model from ΛCDM.

| Model | elpd |
|---|---|
| $f(Q$ as Dark Energy | $-24.47 \pm 6.21$ |
| $\Lambda$CDM | $-25.83 \pm 6.22$ |

Table 6.4: The value of the elpd, computed by PSIS-LOO-CV, for the model presented in eq. (6.1) and $\Lambda$CDM, when considering SnIa from the Pantheon sample.

From the constraining procedure using SnIa from the Pantheon sample, we have obtained the value of $\Omega_m = 0.335 \pm 0.012$ for the model which we are studying in this chapter, and $\Omega_m = 0.285 \pm 0.012$ for $\Lambda$CDM.

If we now remember the constrains set on this model using SS events, which was already presented in table 6.3, we obtained a value for the matter density of $\Omega_m = 0.257 \pm 0.016$ when using the ET. Comparing the constrains set by the SS and SnIa, it is easy to see that there exists a tension between the datasets. In fig. 6.6 we show the corner plot where we have constrained this model (on the left subplot) and $\Lambda$CDM (on the right subplot) with SS events from the ET, LISA (both the median and best catalog) and SnIa alone.

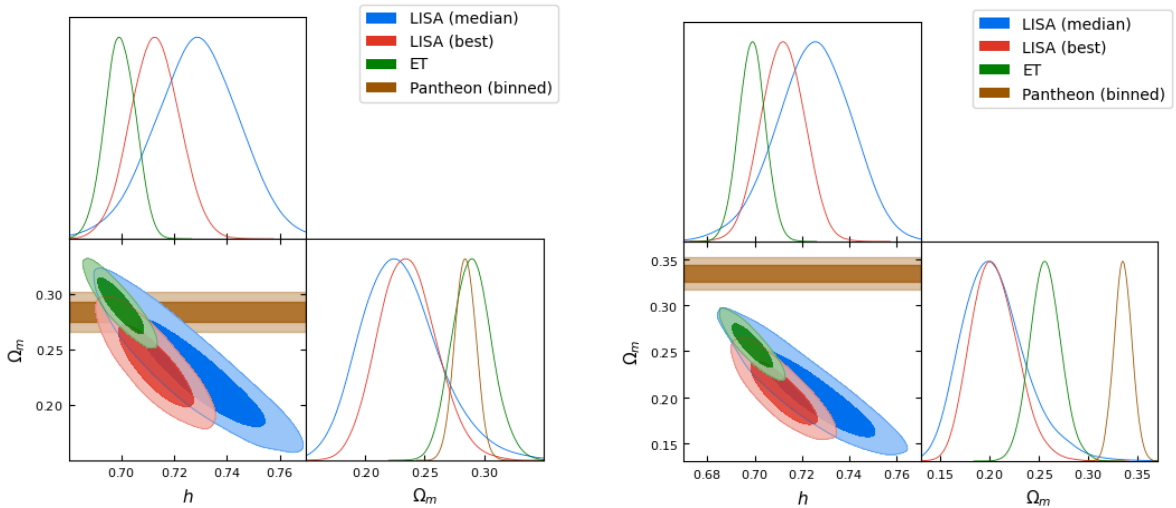

Figure 6.6: Constrains set by the ET, the best and median LISA catalog as well as SnIa from the Pantheon sample alone, for $\Lambda$CDM (on the left) and for the model given in eq. (6.1) (on the right).

From fig. 6.6 it is easy to see that although the events coming from SS are in very good agreement with the data from SnIa in $\Lambda$CDM, the same does not hold true for this model. As such, we can conclude that if $\Lambda$CDM with values of $h = 0.7$ and $\Omega_m = 0.284$ is the model which accurately describes the evolution of the universe, then it is expected that future SS events, along with current SnIa data, will be able to rule this model of $f(Q)$ gravity out. Until then, this model might very well be indistinguishable from $\Lambda$CDM due to its close similarity with it.

To provide a more quantitative view of the existing tension, in table 6.5 we present the number of sigmas between the best-fit of $\Omega_m$ for that catalog and the value of $\Omega_m$ given by the Pantheon sample. This table does not include the worst LISA catalog, due to its remarkably low constraining power when compared to all of the other catalogs.

| Dataset | $\sigma$'s to Pantheon $\Omega_m$ best-fit |
|---|---|
| ET | 4.9 |
| LISA (best) | 5.7 |
| LISA (median) | 4.2 |
| LISA (worst) + LIGO-Virgo (best) | 4.6 |
| LISA (worst) + LIGO-Virgo (median) | 2.0 |
| LISA (worst) + LIGO-Virgo (worst) | 3.2 |

Table 6.5: The number of sigmas between the best-fit value for $\Omega_m$ between each catalog and the Pantheon best-fit, for the model presented in eq. (6.1). The value of $\sigma$ being considered is referent to each dataset.

## 6.3 Summary

In this chapter, we studied an $f(Q)$ model which aims to replace dark energy with a modification of gravity, and introduces no additional free parameters, since the parameter $\lambda$ which was added on the action can be expressed as a function of the densities. This model approaches GR as the redshift increase, meaning that deviations from $\Lambda$CDM only happen in the low redshift regime.

This model was then studied using dynamical system analysis, which reveals three fixed points, of which only one of them is stable, and represents a scenario of permanent inflation. The system is separated into three disjoint regions, which are categorized by the sign of $\lambda$. The region with $\lambda > 0$ is the one that includes the stable fixed point, and the evolution of the universe will therefore always go towards permanent inflation. For the case where $\lambda = 0$, we fall back to STEGR, but not to $\Lambda$CDM, due to the lack of dark energy, meaning that we have a CDM universe. In this region, all trajectories flow from a radiation dominated epoch to a matter dominated epoch. The final region, which features $\lambda > 0$, has no fixed points and all trajectories point towards a big crunch.

We used model selection criteria using SS to attempt to distinguish this model from $\Lambda$CDM apart, only to conclude that no number of SS events is expected to be able to do so. We then decided to increase the number of SS events in each dataset, and see whether applying model selection criteria would allow us to tell this model from $\Lambda$CDM apart. This analysis showed that we do not expect to be able to tell between these two model apart by increasing the number of SS events we expect to measure. Due to the fact that the best-fit for the luminosity distance curves for both models are indistinguishable, we believe that no number of SS events will be able to prefer one model over the other.

Using SnIa data from the Pantheon sample, we showed that both models provide similar best-fits and are equally good at predicting held-out data. However, by constraining this model using SnIa we had a best-fit of $\Omega_m = 0.3353 \pm 0.0091$, whereas when using SS events from the ET we obtained a value of $\Omega_m = 0.257 \pm 0.016$, implying that there is a tension in this model when considering different sources of data. This means that, if our fiducial cosmology is in fact true, then SS are expected to give rise to a tension in this model, and effectively rule it out.

# Chapter 7

# Final Remarks

In this dissertation, we constructed and used SS mock catalogs to evaluate whether future GW observations will be able to distinguish two different cosmological models based on non-metricity, in the form of $f(Q)$ gravity, from $\Lambda$CDM. The mock catalogs were developed for the LIGO-Virgo collaboration, as well as for future ground and space based observatories, LISA and the ET. The mock catalogs were generated assuming a $\Lambda$CDM fiducial model.

The first model studied is the most general $f(Q)$ function that replicates a $\Lambda$CDM background. It introduces only one additional free parameter with respect to $\Lambda$CDM, $\alpha$, and gives rise to an effective gravitational constant such that it modifies the luminosity distance for GWs. Due to a degeneracy between $\Omega_m$ and $\alpha$, we added SnIa data of the Pantheon sample to fix the value of $\Omega_m$. The quality of the constrains for this model seem to increase when LISA is able to measure events at low redshifts. We showed that the ET is expected to be able to set the bounds on the value of $\alpha$ within a $1\sigma$ region of $\sigma_\alpha = 0.25$. The best case scenario for LISA shows a $1\sigma$ region for $\alpha$ which is 0.5 larger when compared to the one expected for the ET. The median LISA catalog is expected to provide a $1\sigma$ region for $\alpha$ which is 2 times larger than that of the ET, while the worst case scenario for LISA is expected to be almost 7 times larger than the ET. While LIGO-Virgo will not be able to constrain this model by itself, it will be able to improve the quality of the constrains set by LISA, which in a best case scenario is expected to have a $1\sigma$ region for $\alpha$ which is 0.8 larger when compared to the ET.

The second model attempts to replace dark energy with a modification of gravity, without introducing additional free parameters, as the parameter which was inserted in the action, denoted $\lambda$, can be expressed as a function of the energy-matter content of the universe. By performing a dynamical system analysis, we showed that this model can either undergo permanent expansion, a big crunch or a CDM type of universe depending on the sign and value of $\lambda$. To match current observations, a value of $\lambda > 0$ is chosen such that the universe expands at late times. We then verified that no number of SS events are able to distinguish between this model and $\Lambda$CDM. The same is true when using SnIa coming from the Pantheon sample. However, it was shown that in this model, there is a tension between SnIa and SS, meaning that, in the future, the data obtained will be able to rule this model out.

Based on this work it is possible to see that SS events are expected to be of great use for observational cosmology. This follows from the fact that the luminosity distance of these events is highly sensitive to modifications of gravity, both from the background and from perturbative effects. Additionally, these events are expected to be measured in a range of redshifts still to be explored. With the remarkable feature of not requiring a ladder distance in order to calibrate the measurements, they bring a new and independent way of testing our cosmological models. Additionally, the different GWs observatories can complement each other by creating a large array of GWs detectors, capturing events taking place throughout the universe.

In the coming future, we aim to constrain the model which was studied in chapter 6 with current data using various observables to ensure its consistency. In particular, we wish to see how this model behaves in the early universe by constraining it with data coming from the Cosmic Microwave Background (CMB). Constraining this model with real data will allow us to make predictions regarding the distribution of SS events and evaluate its differences with respect to $\Lambda$CDM. Once the real data becomes available, we expect to see which one better fits to the measured events (which can be done, for instance, using the posterior predictive distribution). Additionally, we would like to generate a larger number of SS mock catalogs in order to better understand the range of possible catalogs, and to be able to quantify how likely the more extreme scenarios are, as well as which are the most likely outcomes. We also seek to understand and implement better statistical tools in order to quantify the tension that exists between the SnIa and SS in the model presented in chapter 6.